\documentclass[fleqn,usenatbib]{mnras}

\usepackage{amsmath}

\usepackage[T1]{fontenc}

\DeclareRobustCommand{\VAN}[3]{#2}
\let\VANthebibliography\thebibliography
\def\thebibliography{\DeclareRobustCommand{\VAN}[3]{##3}\VANthebibliography}

\usepackage{amssymb}	
\usepackage{graphicx}	
\usepackage{txfonts}
\usepackage{titlesec}
\usepackage{multibib}
\usepackage[dvipsnames]{xcolor}
\usepackage{enumitem}
\usepackage{fancyhdr}
\usepackage[nointegrals]{wasysym} 
\usepackage{subcaption}
\usepackage{orcidlink}

{

\title[\ion{O}{VII} and \ion{O}{VIII} absorption from Hydrangea filaments against cool-core galaxy clusters]{Prospects for detecting the circum- and intergalactic medium in X-ray absorption using the extended intracluster medium as a backlight.}

\author[L. {\v S}tofanov{\' a} et al.]{L{\' y}dia {\v S}tofanov{\' a},$^{1,2}$\thanks{E-mail: stofanova@strw.leidenuniv.nl}\orcidlink{0000-0003-0049-6205}
Aurora Simionescu,$^{2,1,3}$
Nastasha A. Wijers,$^{4}$
Joop Schaye,$^{1}$
Jelle S. Kaastra,$^{2,1}$
\newauthor Yannick M. Bah{\' e},$^{5,1}$ 
and Andr{\' e}s Ar{\' a}mburo-Garc{\' i}a$^{6}$
\\
$^{1}$Leiden Observatory, Leiden University, Niels Bohrweg 2,
2333 CA Leiden, The Netherlands\\
$^{2}$SRON Netherlands Institute for Space Research, Niels Bohrweg 4, 2333 CA Leiden, The Netherlands\\
$^{3}$Kavli Institute for the Physics and Mathematics of the Universe, The University of Tokyo, Kashiwa, Chiba 277-8583, Japan\\
$^4$Center for Interdisciplinary Exploration and Research in Astrophysics (CIERA) and Department of Physics and Astronomy, Northwestern University,\\ ~ 1800 Sherman Avenue, Evanston, IL 60201, USA \\
$^{5}$Institute of Physics, Laboratory of Astrophysics, Ecole Polytechnique Fédérale de Lausanne (EPFL), Observatoire de Sauverny, 1290 Versoix, Switzerland \\
$^{6}$Institute Lorentz, Leiden University, Niels Bohrweg 2, Leiden, NL-2333 CA, the Netherland
}

\date{Accepted 2023 November 14. Received 2023 November 10; in original form 2023 August 10}

\pubyear{2023}

\begin{document}
\label{firstpage}
\pagerange{\pageref{firstpage}--\pageref{lastpage}}
\maketitle

\begin{abstract}

The warm-hot plasma in cosmic web filaments is thought to comprise a large fraction of the gas in the local Universe. So far, the search for this gas has focused on mapping its emission, or detecting its absorption signatures against bright, point-like sources. Future, non-dispersive, high spectral resolution X-ray detectors will, for the first time, enable absorption studies against extended objects. Here, we use the Hydrangea cosmological hydrodynamical simulations to predict the expected properties of intergalactic gas in and around massive galaxy clusters, and investigate the prospects of detecting it in absorption against the bright cores of nearby, massive, relaxed galaxy clusters. We probe a total of $138$ projections from the simulation volumes, finding $16$ directions with a total column density $N_{\ion{O}{VII}} > 10^{14.5}$\,cm$^{-2}$. The strongest absorbers are typically shifted by $\pm 1000$\,km/s with respect to the rest frame of the cluster they are nearest to. Realistic mock observations with future micro-calorimeters, such as the Athena X-ray Integral Field Unit or the proposed Line Emission Mapper (LEM) X-ray probe, show that the detection of cosmic web filaments in \ion{O}{VII} and \ion{O}{VIII} absorption against galaxy cluster cores will be feasible. An \ion{O}{VII} detection with a $5\sigma$ significance can be achieved in $10-250$\,ks  with Athena for most of the galaxy clusters considered. The \ion{O}{VIII} detection becomes feasible only with a spectral resolution of around $1$\,eV, comparable to that envisioned for LEM. 
\end{abstract}

\begin{keywords}
large-scale structure of Universe -- galaxies: clusters: general -- X-rays: galaxies: clusters -- intergalactic medium -- quasars: absorption lines -- local interstellar matter

\end{keywords}



\section{Introduction}
\label{Sec:intro}
The matter distribution in the Universe, as we know and understand it today, is believed to have a filamentary structure on its largest scales -- famously known as the cosmic web \citep{1986ApJ...302L...1D, 1996Natur.380..603B}. The cosmic web formed when the initial small density field fluctuations collapsed under the influence of gravity, which resulted in structures such as cosmic sheets, voids, nodes and filaments. In this paper we to focus on filaments, which are the bridges between the cosmic nodes. In the nodes, which are the places with the highest concentration of dark matter, galaxy clusters form through hierarchical structure growth. 

Cosmological simulations predict that at $z \approx 0$, most of the gas resides in the warm-hot intergalactic medium (WHIM), which contains around $30$\%, and possibly up to $60$\% of all the baryons in the Universe (e.g. \citealp{1999ApJ...514....1C, 2001ApJ...552..473D, 2012MNRAS.425.1640T, 2019MNRAS.486.3766M, 2021A&A...646A.156T}). According to the simulations, this diffuse WHIM has been shock-heated to temperatures of $10^5 - 10^7$\,K. It has low electron densities, $n_\textnormal{e} \sim 10^{-6} - 10^{-4}$\,cm$^{-3}$ (only $10-10^3$ times the mean baryon density of the Universe), which makes it extremely difficult to observe in emission due to its low surface brightness. The high temperatures make \ion{O}{VI} ($10^{5.3} < T < 10^{5.8}$\,K, in collisional ionisation equilibrium - CIE) together with \ion{O}{VII} ($10^{5.4} < T < 10^{6.5}$\,K, in CIE) and \ion{O}{VIII} ($10^{6.1} < T < 10^{6.8}$\,K, in CIE) good tracers of the WHIM.

Multiple papers have reported detections of the WHIM in X-ray emission (e.g. \citealp{ 1999A&A...341...23K,  2003A&A...410..777F, 2003A&A...397..445K, 2008A&A...482L..29W, 2015Natur.528..105E, 2017A&A...606A...1A, 2018ApJ...867...25C, 2021A&A...647A...2R}). However, since the X-ray emissivity increases as the density squared, extreme ultraviolet (EUV) and X-ray emission tends to probe the densest and hottest parts of the WHIM (e.g. \citealp{2008SSRv..134..295B, 2010MNRAS.407..544B, 2010MNRAS.408.1120B, 2013MNRAS.430.2688V, 2022MNRAS.514.5214W, 2023MNRAS.523.1209C}).

Because absorption scales linearly with the density, it is less biased to higher density gas than emission, making absorption more suitable for detecting the bulk of the WHIM (see e.g. \citealp{2019MNRAS.488.2947W, 2020MNRAS.498..574W}). A detection of the WHIM in \ion{O}{VII}, \ion{O}{VIII}, and \ion{Ne}{IX} absorption toward quasar PKS 2155-304 was reported by \citet{2002ApJ...573..157N}. Later detections were reported by \citet{2003ApJ...586L..49F} and \citet{2003ASSL..281..109R} in \ion{O}{VII}, and \citet{2004PASJ...56L..29F} in \ion{O}{VIII}. A detection of two filaments towards the sightline of the blazar Mrk $421$ with Chandra was documented in \citet{2005ApJ...629..700N}, however, later \citet{2006ApJ...652..189K} re-analysed the same dataset and found the observations of \citet{2005ApJ...629..700N} were statistically insignificant. \citet{2007ApJ...655..831T} claimed a detection of WHIM in \ion{Ne}{IX} and \ion{O}{VIII}. Considering more recent works,  \citet{2018Natur.558..406N} found two absorbers toward blazar 1ES 1553+113 with reported \ion{O}{VII} column densities of a few times $10^{15}$\,cm$^{-2}$. The same sightline was also explored by \citet{2023arXiv230501587S} who reported a $\sim 3\sigma$ detection in \ion{O}{VII}. \citet{2021A&A...656A.107A} detected \ion{O}{VI} and \ion{O}{VII} absorptions toward the galaxy Ton S 180 with a significance higher than $5\sigma$ and reported column densities of few times $10^{14}$\,cm$^{-2}$ and $10^{16}$\,cm$^{-2}$ for \ion{O}{VI} and \ion{O}{VII}, respectively, while the \ion{O}{VIII} detection was not significant. We emphasize, however, that most of the absorption studies performed with current missions are of a low statistical significance and often challenging in terms of the systematic uncertainties, and therefore the robustness of these detections often remains debated in the literature. \citet{2022arXiv220315666N} summarizes the history of WHIM absorption studies towards bright point-like sources, and discusses the prospects of future WHIM detections.

A complementary method to observe gas in filaments in absorption is to use galaxy clusters as background candles \citep{1999ApJ...522L..13M, 2009astro2010S.192M}. Since galaxy clusters are located at the intersection of cosmic web filaments, sightlines pointed towards cluster cores are likely to probe the densest parts of the surrounding filaments, and therefore provide the highest column densities and hence the strongest absorption profiles. \citet{2021ExA....51.1043S} extended this idea by providing estimates of the absorption from a simplified WHIM component, approximated as a single temperature, single density, and single metallicity gas. Here we explore this idea further, and provide predictions for multiple cosmic web filaments taken from cosmological hydrodynamical simulations. We simulate observations of these filaments with the future mission Athena \citep{2018SPIE10699E..1GB} and the proposed X-ray probe Line Emission Mapper (LEM, \citealp{2022arXiv221109827K}), while taking into account a more complex (three component) model of the Galactic interstellar medium (ISM).


Since we are interested in large-scale structure filaments connecting to galaxy clusters, we need to use cosmological simulations covering a large volume, in order to contain enough massive haloes for our study. On the other hand, the properties of the WHIM depend sensitively on the physics of galaxy evolution, including processes like feedback and star formation, for which it is important to resolve small spatial scales. Cosmological zoom-in simulations are therefore well suited for this study. Specifically, we focus on the Hydrangea simulation suite \citep{2017MNRAS.470.4186B}, containing a sample of $24$ galaxy clusters identified from dark matter only simulations, and then re-simulated in a volume up to $10 \times r_{200c}$\footnote{$r_{200c}$ denotes the radius of the sphere within which the mean overdensity is $200$ times the critical density of the Universe.} including baryons and using the same resolution and galaxy formation model as the EAGLE simulations (’Evolution and Assembly of GaLaxies and their Environments’, \citealp{2015MNRAS.446..521S, 2015MNRAS.450.1937C}). Additional details about these simulations can be found in Sec.\,\ref{Sec:Hydrangea}. Hydrangea is specifically tailored towards describing massive galaxy clusters, $M_{200, \rm c}\in(10^{14} - 10^{15.4})$\,M$_{\astrosun}$, which may make the cosmic web filaments connected to these clusters denser and therefore easier to detect.



\subsection{Research set-up}
\label{sec:intro:readme}

The main goal of this paper is to determine whether extended sources of light, in particular galaxy clusters, can serve as background sources for detection of the cosmic web filaments in \ion{O}{VII} or \ion{O}{VIII} absorption. We use the cosmic web filaments from the Hydrangea cosmological hydrodynamical simulations and calculate their properties by probing the gas with many lines of sight (LoS). We create column density maps and average optical depth profiles for \ion{O}{VII} and \ion{O}{VIII}.
 

As background sources we select observed bright, relaxed cool-core galaxy clusters, which span a range of redshifts, masses, and temperatures. Although our method can in principle be used with any extended source, these galaxy clusters should be among the most promising background sources for our study. We calculate the spectral energy distributions (SEDs) and the surface brightness profiles (SB) of these galaxy clusters. The SB profiles are used to weight the spatially dependent optical depth profiles from the simulations, while the SEDs are used as the backlight sources. 

To realistically simulate the detectability of the absorption from the cosmic web filaments, we use a three-component model of the Milky Way absorption. We simulate the resulting model spectra using response files for realistic X-ray detectors and we determine the significance of the \ion{O}{VII} and \ion{O}{VIII} detection from the cosmic web filaments for various exposure times. 


This paper is organised as follows. In Sec.\ref{Sec:methods} we summarize our methodology (Sec.\,\ref{Sec:methods_simulations}) and results (Sec.\,\ref{Sec:results}) for the Hydrangea cosmological simulations. In Sec.\,\ref{Sec:observational_predictions_section} we summarize our methodology (Sec.\,\ref{sec:observations_methods}) and results (Sec.\,\ref{Sec:XIFU_results}) for the observational predictions. In Sec.\,\ref{Sec:discussion} we discuss the effects of choosing a different size of the radius for which we extracted the absorption profiles from the simulations, the impact of different SB weighting in the absorption profiles, and how different detectors or an additional radiative cooling in cluster cores might affect our results. And lastly, we provide our main conclusions in Sec.\,\ref{Sec:conclusions}.

\section{Cosmological simulations}
\label{Sec:methods}

\subsection{Methods}
\label{Sec:methods_simulations}
\subsubsection{Hydrangea simulations}
\label{Sec:Hydrangea}

In this section we summarize the main properties of the Hydrangea simulations. For a more detailed description we refer to \citet{2017MNRAS.470.4186B} and \citet{2015MNRAS.446..521S}. Hydrangea is also part of the Cluster-EAGLE (C-EAGLE, CE) project described in \citet{2017MNRAS.471.1088B}.

The Hydrangea simulations are cosmological hydrodynamical zoom-in simulations that use the EAGLE galaxy formation model (\citealp{2015MNRAS.446..521S, 2015MNRAS.450.1937C}). They use a modified version of the $\mathrm{GADGET 3}$ smoothed particle hydrodynamics (SPH) code last described by \citet{2005MNRAS.364.1105S}. The set of modifications to the hydrodynamics and the time-stepping scheme of $\mathrm{GADGET 3}$, known as 'Anarchy', consists of removing the unphysical surface tension at contact discontinuities by using the discrete particle Lagrangian SPH formulation from \citet{2013MNRAS.428.2840H}; the artificial viscosity switch from \citet{2010MNRAS.408..669C}; a switch for artificial thermal conductivity \citep{2008JCoPh.22710040P}; the $C^2$ Wendland kernel \citep{Wendland1995}; and the energy conserving time-step limiters from \citet{2012MNRAS.419..465D}. These modifications are described in more detail in appendix A of \citet{2015MNRAS.446..521S}, and in \citet{2015MNRAS.454.2277S}. The cosmological parameters were adopted from \citet{2014A&A...571A..16P}.

Hydrangea implements a number of astrophysical processes through sub-resolution (sub-grid) models. The radiative cooling and photoheating rates are based on \citet{2009MNRAS.393...99W} which assumes a \citet{2001cghr.confE..64H} ionizing UV and X-ray background. The star formation rate of gas particles follows the pressure law given by \citet{2008MNRAS.383.1210S}. The mass and metal enrichment of gas due to stellar mass loss is described in \citet{2009MNRAS.399..574W} and \citet{2015MNRAS.446..521S}. The star formation energy feedback happens in a single thermal mode, with a small number of gas particles heated by a large temperature difference \citep{2012MNRAS.426..140D}. The energy feedback from supermassive black holes ('AGN feedback') is implemented in a similar fashion \citep{2009MNRAS.398...53B}. 
 
The high-resolution zoom-in regions of Hydrangea have the same baryonic mass resolution, $m_{\rm baryon} = 1.81 \times 10^6$\,M$_{\astrosun}$, as the largest EAGLE volume and the gravitational softening length is $\epsilon = 0.7$\,pkpc (physical kpc) at redshift ($z<2.8$). The zoom-in regions were created by running dark matter only simulations with the $\mathrm{GADGET}$  code but for a much larger volume (3200 co-moving Mpc)$^3$. The cluster candidates were selected from the parent simulation at redshift $z=0$ based on three criteria:
\begin{itemize}[leftmargin=5pt]
	\item $M_{200 \rm c} \geq 10^{14}$\,M$_{\astrosun}$.
	\item There is no other massive halo located within $30$\,pMpc or within $20 \times r_{200 \rm c}$ (the larger of these two was applied).
	\item No galaxy cluster candidate for re-simulation is located closer than $200$\,pMpc from any of the periodic simulation box edges. 
\end{itemize} 

Application of these three criteria resulted in $30$ galaxy cluster candidates that were re-simulated in a high-resolution zoom-in box while including baryons. Outside of the zoom-in region, the box is filled with low-resolution particles that interact only gravitationally. We used $23$ clusters at redshift $z=0$, for which the zoom-in region was $10 \times r_{200 \rm c}$, which is big enough to simulate the intra-cluster medium and the filamentary structures in the vicinity of the galaxy clusters.  Six clusters were excluded because they were simulated only to $5 \times r_{200 \rm c}$, and CE-$10$ was excluded because it had unphysically powerful AGN explosions due to a bug in the code, and therefore the gas properties were not considered reliable.

\subsubsection{Specwizard}
\label{Sec:specwizard}

In order to calculate \ion{O}{VII} and \ion{O}{VIII} column densities in a LoS through the Hydrangea zoom-in volumes, we use the software \textit{Specwizard}, which creates the synthetic spectra from the smoothed particle hydrodynamics simulations. In order to create these synthetic spectra, \textit{Specwizard} uses the physical properties of the simulated gas particles (e.g. density, temperature, metal abundances), and accounts for photoionization from the UV and X-ray background. As shown in \citet{2022MNRAS.515.3162S}, in certain cases, photoionization from galaxy cluster photons can also play a role. However, for the case of large-scale filaments aligned predominantly along the line of sight, this effect was found to be negligible; therefore, we do not include it in the present work.


\textit{Specwizard} was developed by Joop Schaye, Craig M. Booth and Tom Theuns (see Sec.\,3.1 of \citealp{2011MNRAS.413..190T}). In our study we used its python version, which is described in more detail in  Arámburo-García+2023 (in prep).

\subsection{Results: properties of Hydrangea absorbers}
\label{Sec:results}
\begin{figure}
	\centering
	\includegraphics[width=0.48\textwidth]{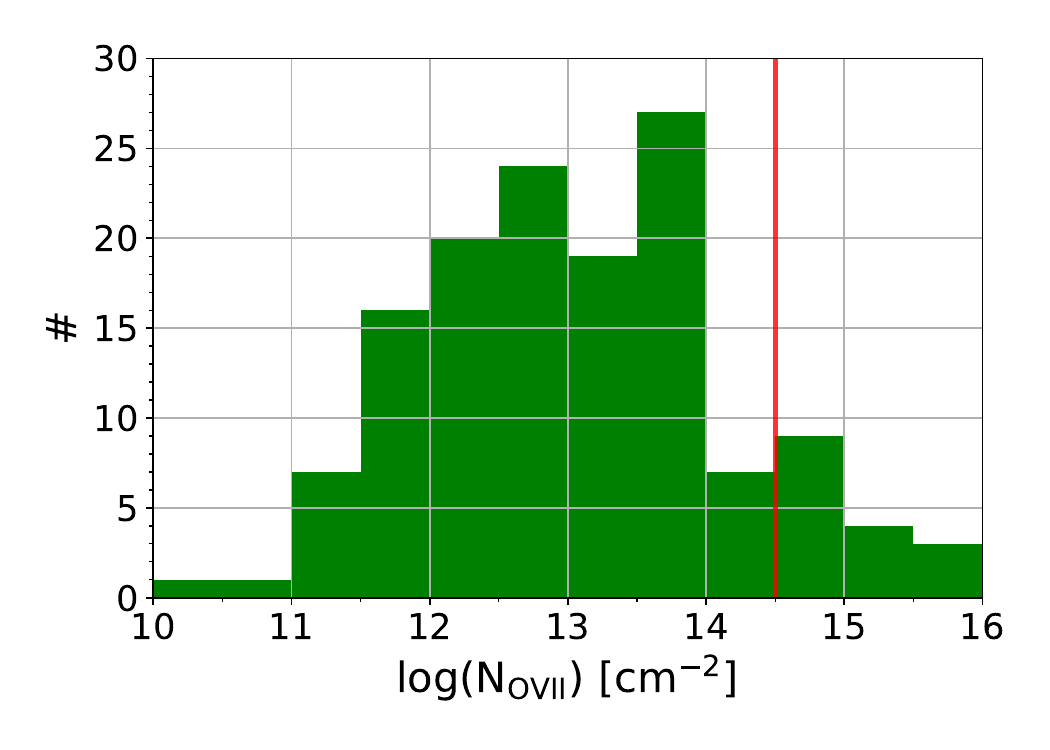} 
	\caption{Histogram of the number of sightlines (out of the total number of $138$ through Hydrangea cluster centres) as a function of \ion{O}{VII} total column density per line of sight. The width of the column density bins is $0.5$\,dex. The red solid line represents the cut for all LoSs with \ion{O}{VII} total column density higher than $10^{14.5}$\,cm$^{-2}$. This cut selects $16$ directions of lines of sight that are later discussed in the paper. }
	
	\label{Fig:all_absorbers_histogram}
\end{figure}

To get an initial estimate of the absorber statistics, we first examine $23$ clusters in the Hydrangea simulations in $6$ different directions along the $\pm x$, $\pm y$ and $\pm z$ axes of the simulation volumes. The length of each LoS in every direction is the whole zoom-in region of $10 \times r_{200,c}$ (which also includes half of the galaxy cluster).  In Fig.\;\ref{Fig:all_absorbers_histogram} we show the results of this initial inspection, where we draw only one LoS per every direction pointed at the cluster centre, and plot a histogram of the total \ion{O}{VII} column density per LoS. We see that the bulk of the LoSs lie between $10^{11.5}$\,cm$^{-2}$ and $10^{14}$\,cm$^{-2}$. From the total number of $138$ different LoSs, we find $16$ (7) LoSs in which the total \ion{O}{VII} column density is higher than $10^{14.5}$\,cm$^{-2}$ ($10^{15}$\,cm$^{-2}$). The observational threshold for \ion{O}{VII} column densities is currently around a few times $10^{15}$\,cm$^{-2}$, with rare exceptions reaching down to a few times $10^{14}$\,cm$^{-2}$. With upcoming future missions this threshold could be pushed to $10^{14}$\,cm$^{-2}$ \citep{2022arXiv220315666N}. Since Hydrangea was designed to study massive galaxy clusters with masses M$_{200, \rm c}$ between $10^{14}$\,M$_{\astrosun}$ and $10^{15.4}$\,M$_{\astrosun}$, it is natural to expect that our studies will be biased towards more massive filaments and therefore higher \ion{O}{VII} (and \ion{O}{VIII}) column densities. Due to the patchy nature of the gas in the cosmic web filaments, the LoS pointed towards the cluster centre does not have to be fully representative of the column densities in the selected directions. Therefore, we also create \ion{O}{VII} column density maps for all $138$ directions (see Sec.\,\ref{Sec:column_density_maps}).  



\subsubsection{\ion{O}{VII} column density maps}
\label{Sec:column_density_maps}

We expand our search for projections where the total \ion{O}{VII} column density per LoS is higher than $10^{14.5}$\,cm$^{-2}$ by creating column density maps for all $138$ possibilities which cover a region with radius of $300$\,kpc around the galaxy cluster centre using $2$\,kpc pixels. With this we confirm that no other directions were found in addition to the $16$ Hydrangea absorbers shown in Fig.\,\ref{Fig:all_absorbers_histogram}. 

\begin{figure}
	\centering
	\includegraphics[width=\columnwidth]{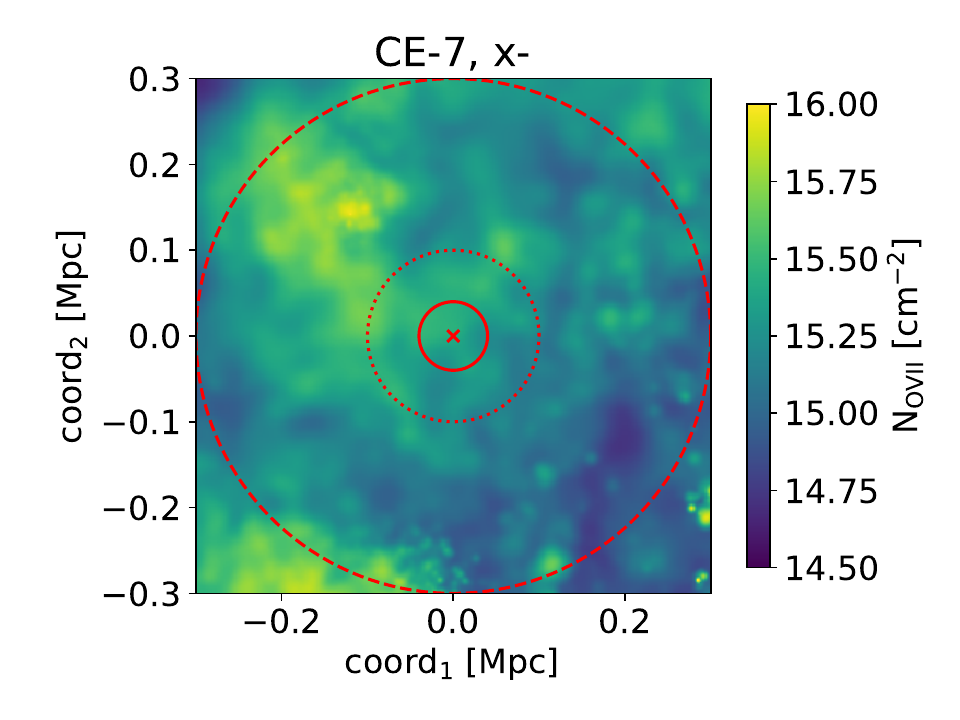} 
	\includegraphics[width=\columnwidth]{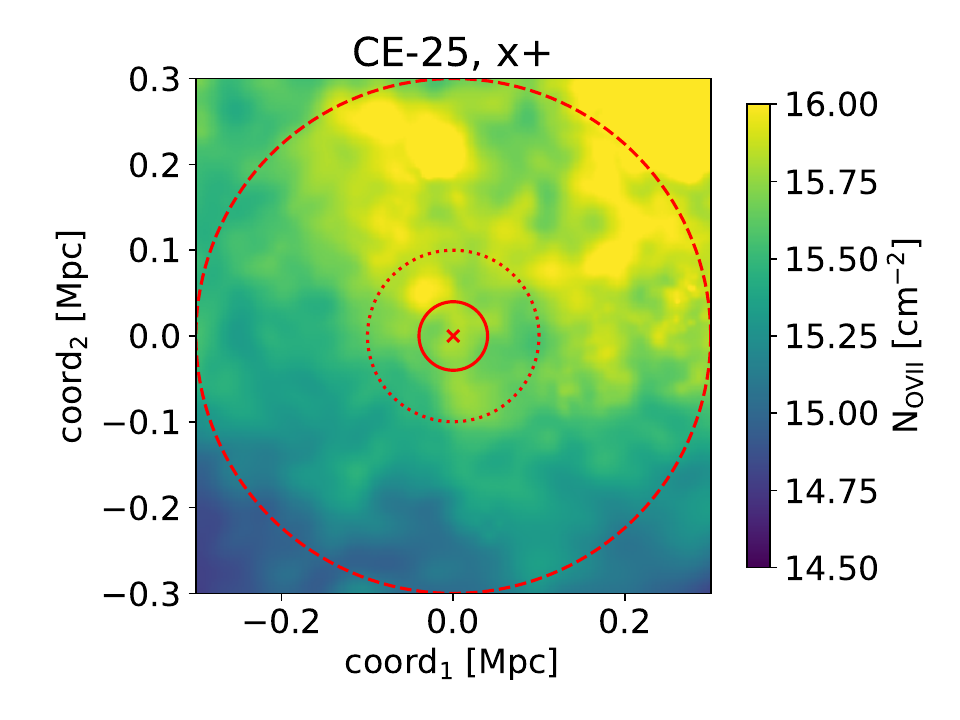} \\
	\includegraphics[width=\columnwidth]{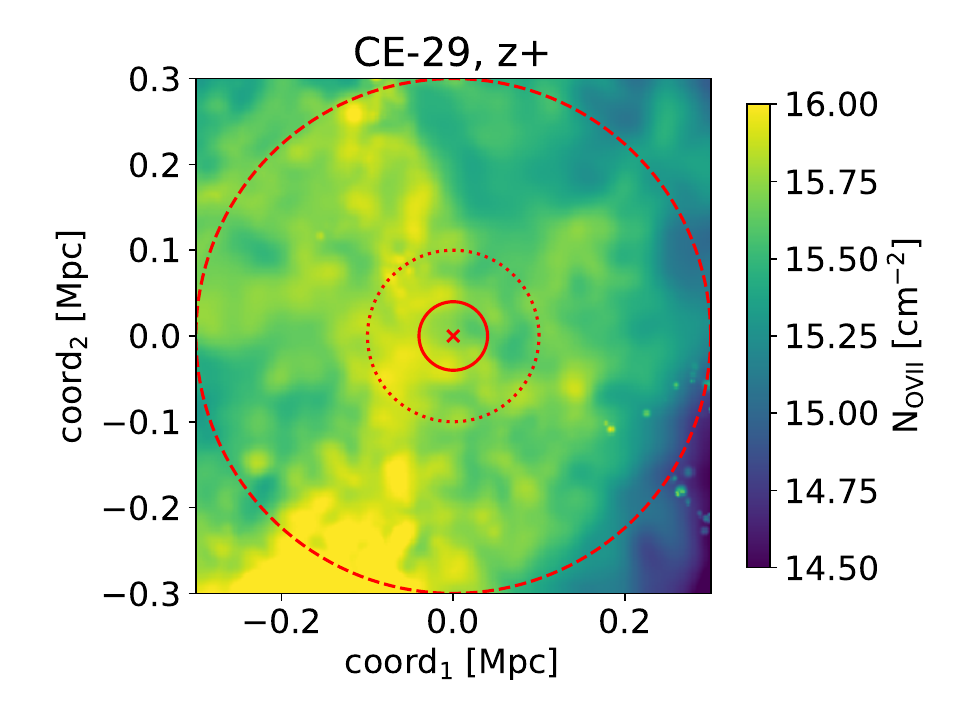} 
	\caption{\ion{O}{VII} column density maps that cover the central region with radius $300$\,kpc around CE-$7$ $x-$, CE-$25$ $x+$ and CE-$29$ $z+$ galaxy clusters. The maps have $300 \times 300$ pixels with a pixel size of $2$\,kpc. We only plot gas with N$_{\ion{O}{VII}} \geq 10^{14.5}$\,cm$^{-2}$. The red circles show  $40$\,kpc (solid line), $100$\,kpc (dotted line) and $300$\,kpc (dashed line) regions around the galaxy cluster centre (red cross).}	
	\label{Fig:column_density_maps}
\end{figure}

In Fig.\,\ref{Fig:column_density_maps} we show an example of three different orientations for three different clusters with halo identifiers: CE-$7$ $x-$, CE-$25$ $x+$ and CE-$29$ $z+$. In Sec.\,\ref{Sec:appendix_column_density_maps} and Fig.\,\ref{Fig:column_density_maps_app1} and Fig.\,\ref{Fig:column_density_maps_app2} we show column density maps of the remaining $13$ directions. In the maps we exclude gas within the central galaxy cluster halo ($r<r_{200, \rm c}$) and for plotting purposes we impose a lower threshold on the column densities of N$_{\ion{O}{VII}} \geq 10^{14.5}$\,cm$^{-2}$. To create these column density maps, we exclude the star forming gas by removing all SPH particles that fall on the equation of state (see Sec.\,4.3 in \citealt{2015MNRAS.446..521S} for more details). We also exclude the gas that was directly heated by supernovae and AGN feedback \citep{2019MNRAS.488.2947W}.


\subsubsection{Spatial properties of the \ion{O}{VII} and \ion{O}{VIII} absorbers}
\label{Sec:spatial_properties}

Next, we create a Cartesian grid of $20\times20$ equidistantly spaced LoSs and select those that are within $R < 100$\,kpc of the projected centre of the cluster. This yields a total number of $276$ LoSs, $10$\,kpc apart, which we used to probe the absorbing gas. We verified that this number is sufficient in order to probe the spatial substructures of the gas in Hydrangea volumes. Decreasing the number of LoSs to $200$, $100$, and $50$ (in comparison to $276$ LOSs), results to a difference of $<3$\%, $<6$\%, and $<8$\%, respectively. 

We average all $276$ LoSs for each of the $16$ directions assuming each LoS adds to the final profile with an equal weight. This allows us to study the \ion{O}{VII} and \ion{O}{VIII} absorbers independently from the exact surface brightness of the extended background source. For the \ion{O}{VII} absorption we take into account only the resonant line at $21.60169$\,$\AA$ with the oscillator strength $0.696$ and the Einstein coefficient of $3.32 \times 10^{12}$\,s$^{-1}$. Since for the absorption studies the lines with the highest Einstein coefficients contribute the most to the absorption line profile, taking into account only the resonant line of \ion{O}{VII} is sufficient. However, for \ion{O}{VIII} there are two lines that have similarly strong Einstein coefficients: (a) the line at $18.96711$\,$\AA$ with oscillator strength $0.2771$ and Einstein coefficient $2.569 \times 10^{12}$\,s$^{-1}$, and (b) the line at $18.97251$\,$\AA$ with oscillator strength $0.1385$ and Einstein coefficient $2.567 \times 10^{12}$\,s$^{-1}$. Therefore, we combine these lines to calculate the \ion{O}{VIII} absorption. The final \ion{O}{VIII} wavelength of $18.96891$\,$\AA$ is obtained as the average of the two strongest lines weighted by their oscillator strengths. The combined oscillator strength is calculated as a sum of both oscillator strengths and results in the final value of $0.4156$. Finally, the Einstein coefficient for the combined \ion{O}{VIII} line is the larger one out of both lines: $2.569 \times 10^{12}$\,s$^{-1}$. This step of combining two lines into one is done just because the \textit{Specwizard} code itself can only take into account (in its short spectra mode) one line at a time.  Because the X-ray detectors considered here have a spectral resolution of $2.5$\,eV (Athena X-IFU, see Sec.\,\ref{Sec:instruments}) and $0.9$\,eV (LEM, see Sec.\,\ref{Sec:discussion_detectors}), this approximation does not affect our main conclusions (the wavelengths of these two lines are separated by $\Delta\lambda = 0.0054$\,$\AA = 0.02296$\,eV.)

\begin{figure}
	\centering
	\includegraphics[width=\columnwidth]{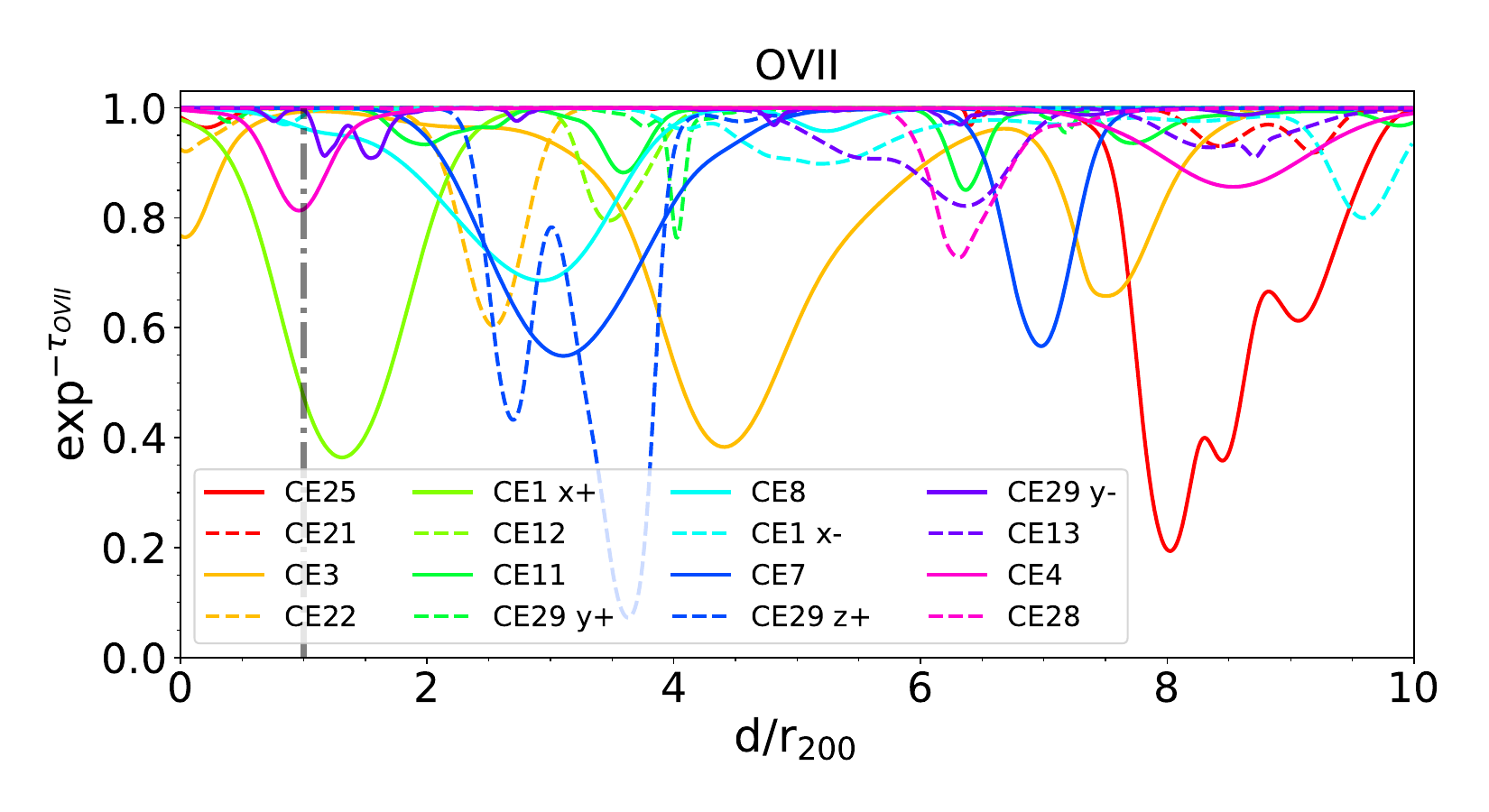} \\
	\includegraphics[width=\columnwidth]{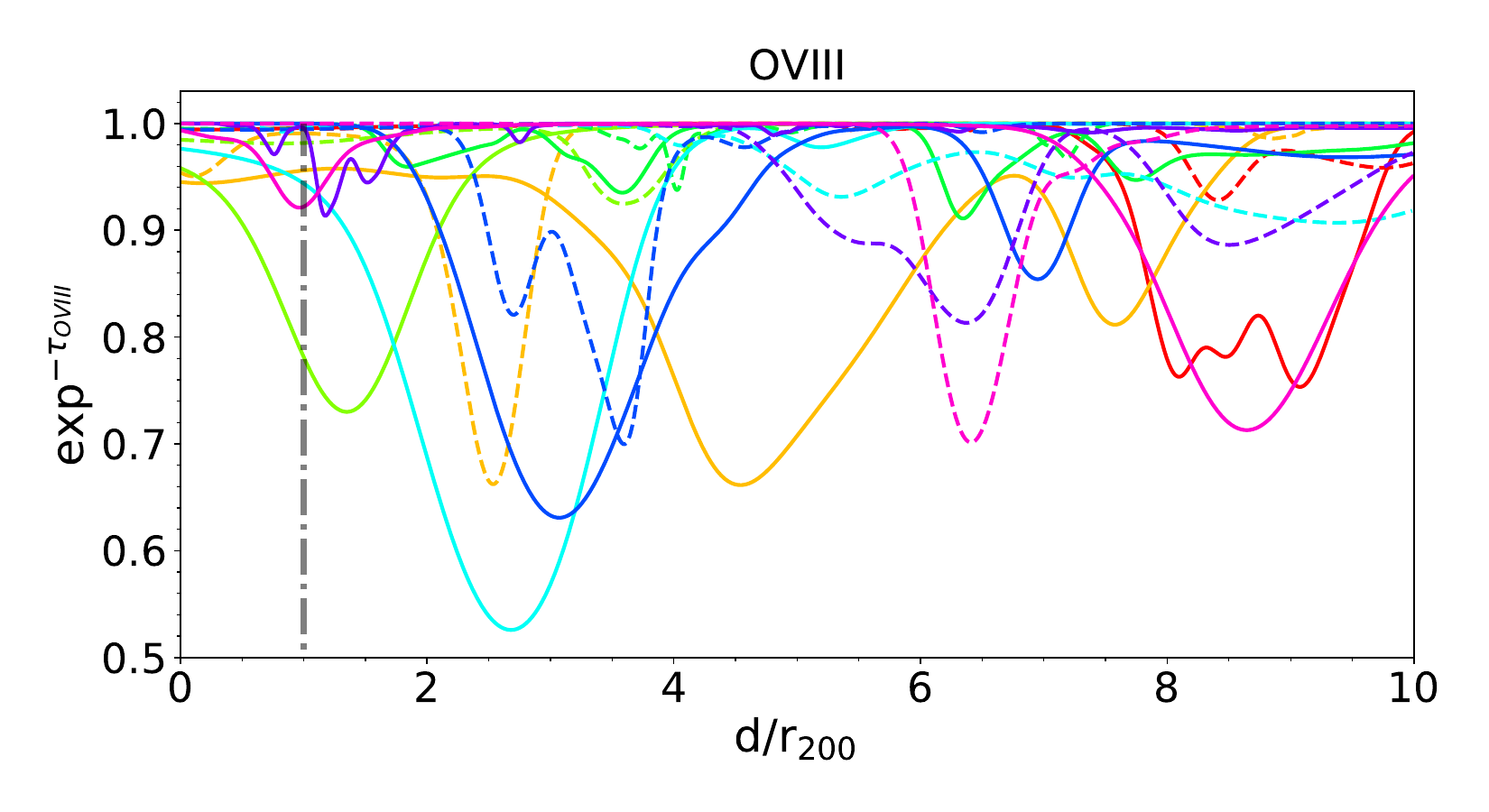}
	\caption{Equally weighted average \ion{O}{VII} (top panel) and \ion{O}{VIII} (bottom panel) absorption profiles within the radius $100$\,kpc from the cluster centre as a function of the position along the LoS $d$. Different colours and line styles represent different projections and Hydrangea clusters. The grey dash-dotted vertical line shows the position of $r_{200}$. For clusters where multiple projections had N$_{\ion{O}{VII}} \geq 10^{14.5}$\,cm$^{-2}$, we specify the projection ($\pm x, \pm y, \pm z$) in addition to the halo identifier. Otherwise, only the halo identifier is shown. We see that the sizes of the most prominent absorbers (as e.g. CE-$25$ or CE-$29$ $z+$) along the line of sight are on average $\sim 2\times r_{200}$. Some cases, e.g. CE-$3$ or CE-$7$, have two prominent absorption lines along the LoS.}
	\label{Fig:tau_vs_position}
\end{figure}

In Fig.\,\ref{Fig:tau_vs_position} we show for all $16$ cases the flux depletion in \ion{O}{VII} and in \ion{O}{VIII} as a function of the position along the LoS $d$, shown as a fraction of $r_{200}$. For clusters where multiple projections had N$_{\ion{O}{VII}} \geq 10^{14.5}$\,cm$^{-2}$, we specify the projection ($\pm x, \pm y, \pm z$) in addition to the halo identifier. Otherwise, only the halo identifier is shown. The four most prominent absorptions are for clusters CE-$29$ $z+$, CE-$25$, CE-$1$ $x+$ and CE-$3$, where the absorption in \ion{O}{VII} is more than $50$\%. In almost all $16$ cases there is more than just one absorption line per direction. The sizes of the most prominent absorbers (e.g. CE-$25$ or CE-$29$ $z+$) along the line of sight are on average $\sim 2\times r_{200}$. Some absorbers as e.g. CE-$3$ or CE-$7$ have two prominent lines along the LoS. The same holds true for \ion{O}{VIII} absorption, however, the overall depth of the absorption features does not exceed $50$\% and the profiles are shallower in comparison with \ion{O}{VII}. This is expected since the cosmic web filaments are not hot enough to ionise a dominant fraction of oxygen to its \ion{O}{VIII} state. 


\subsubsection{Addition of ion-weighted peculiar velocities}
\label{Sec:addition_vpec}

An important aspect of the detectability of filaments in addition to their density, temperature, metallicity, and the total column density in the line of sight, is also how redshifted these absorption lines are in comparison with absorption lines from the ISM of the Milky Way as well as the emission lines from the galaxy cluster itself (relevant mainly for \ion{O}{VIII} because of its brightness in the cluster centre). The velocity difference in the LoS between the gas in the galaxy cluster and the gas in filaments can be calculated by taking into account the ion-weighted peculiar velocities computed by \textit{Specwizard} as well as the peculiar velocity of the galaxy cluster. We define the galaxy cluster peculiar velocity as the centre-of-mass velocity of the central subhalo\footnote{We define the centre-of-mass velocity of the central subhalo as the zero-momentum-frame velocity of the subhalo with index $0$.}.

\begin{figure}
	\centering
	\includegraphics[width=\columnwidth]{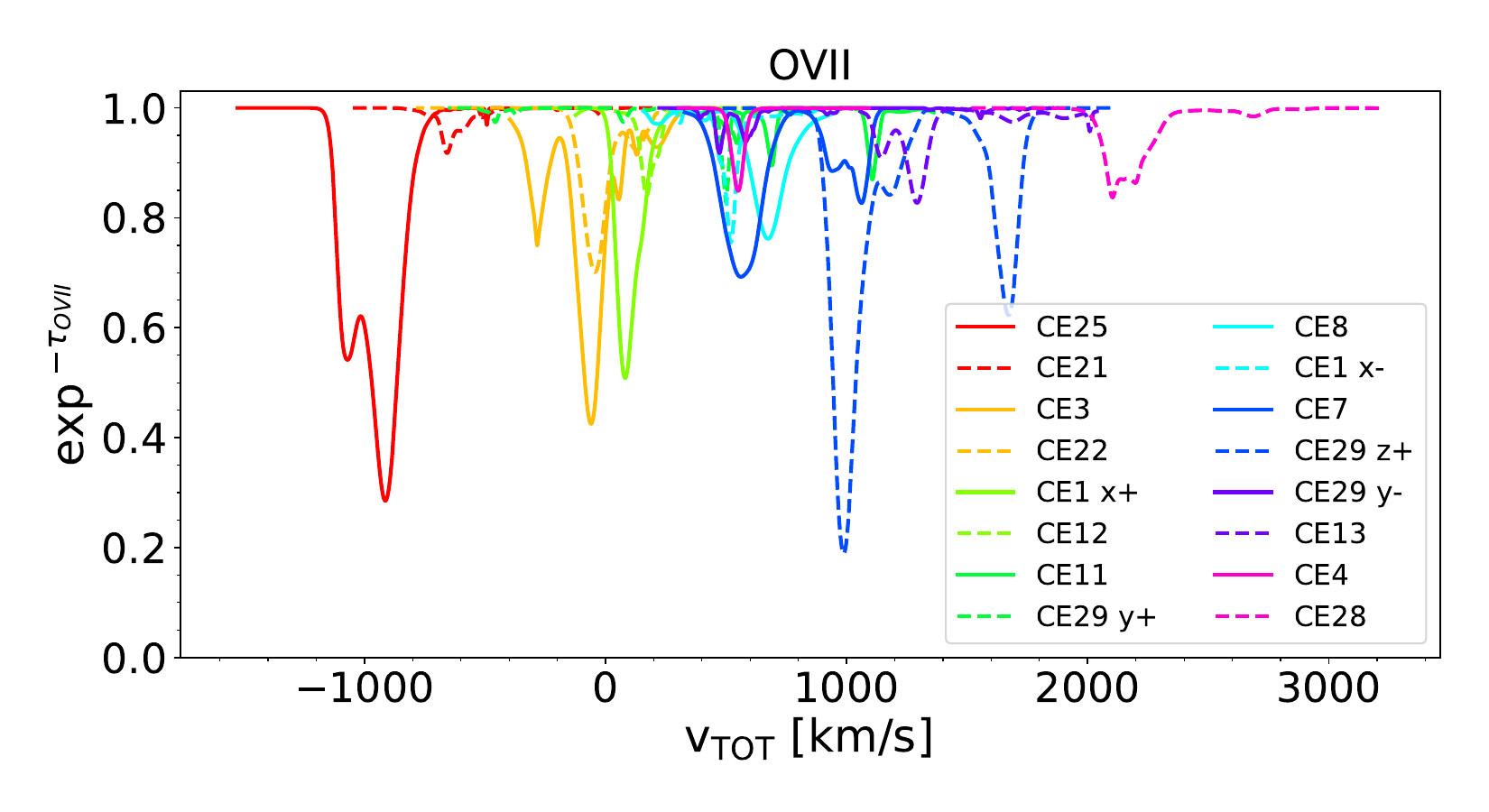} \\
	\includegraphics[width=\columnwidth]{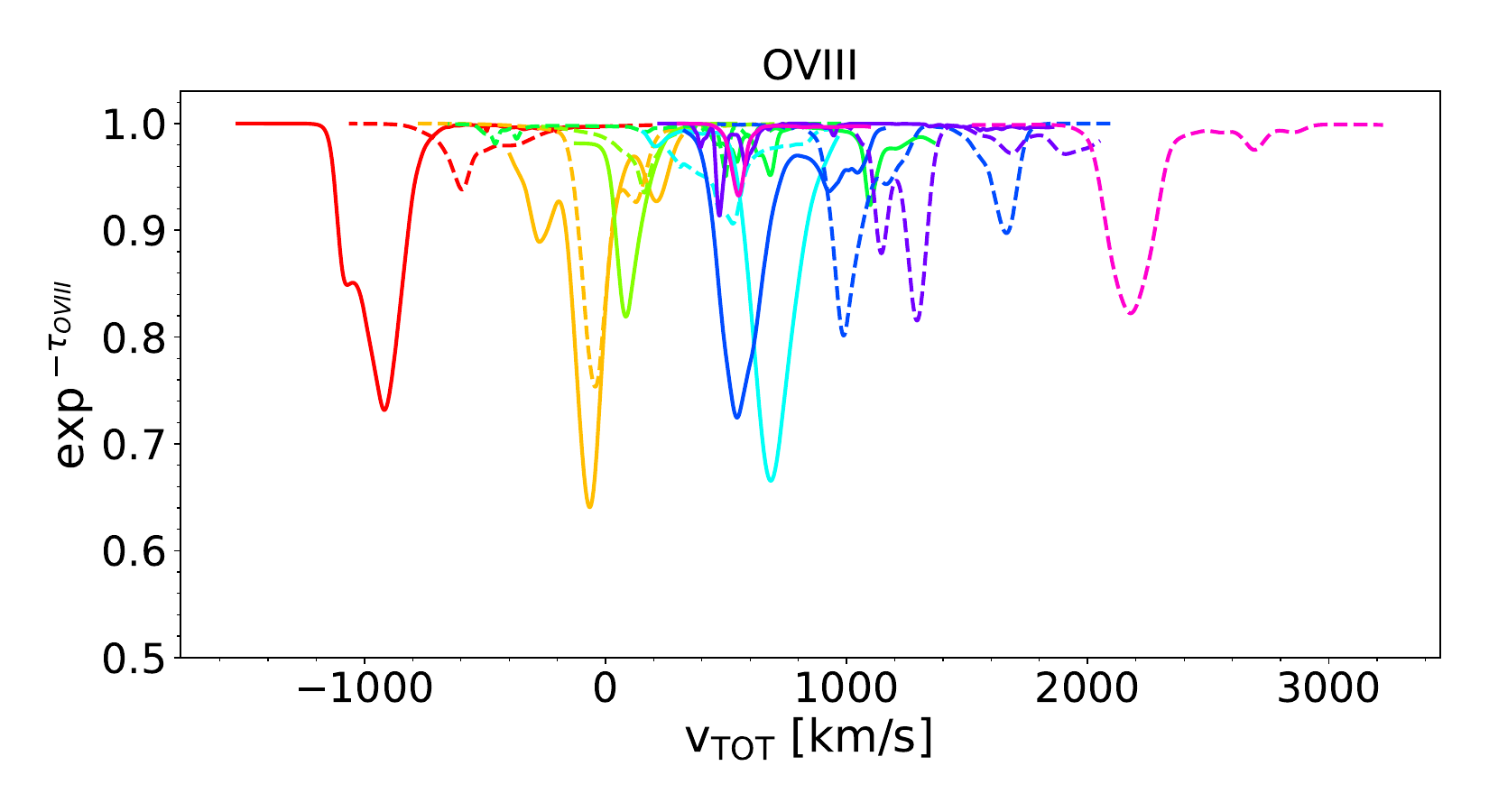}
	\caption{\ion{O}{VII} (top panel) and \ion{O}{VIII} (bottom panel) average absorption profiles (equally weighted) as a function of the total velocity along the line of sight $v_{\rm TOT}$, which is the sum of the Hubble flow velocity and the peculiar velocity of the individual SPH particles that contribute to LoSs. The peculiar velocity of the galaxy cluster is subtracted from the total velocity $v_{\rm TOT}$. Different colours and line styles represent different projections and Hydrangea clusters. The velocity difference between the strongest absorption lines and the galaxy cluster centre is on average $\pm 1000$\,km/s. }
	
	\label{Fig:tau_vs_vtot_vs_energy}
\end{figure}

Fig.\,\ref{Fig:tau_vs_vtot_vs_energy} shows the flux depletion in \ion{O}{VII} and in \ion{O}{VIII} as a function of the total velocity $v_{\rm TOT}$ along the line of sight, which is the sum of the Hubble flow velocity and the peculiar velocity of the individual SPH particles that contribute to LoSs. The peculiar velocity of the galaxy cluster is subtracted from the total velocity $v_{\rm TOT}$. These profiles were calculated by taking an arithmetic average of the full sample of $276$ LoSs for each direction. The total velocities of \ion{O}{VII} and \ion{O}{VIII} absorbers with respect to the cluster range from $-1000$\,km/s to approximately $2000$\,km/s. For most of the \ion{O}{VII} absorbers the addition of the peculiar velocities mainly results in a different shape of the absorption features compared to the profiles shown in Fig.\,\ref{Fig:tau_vs_position}, and it also slightly affects the overall depth of the absorption lines. The addition of peculiar velocities seems to affect the strength of the absorption lines more significantly for \ion{O}{VIII}. 

\section{Observational predictions}
\label{Sec:observational_predictions_section}
\subsection{Methods}
\label{sec:observations_methods}

\subsubsection{Galaxy cluster sample selection}
\label{Sec:vikhlinin_clusters}

For a detection of absorption originating from the cosmic web filaments against extended background sources like galaxy clusters, these background galaxy clusters need to be very bright in their cores. This makes cool-core (CC) clusters perfect candidates due to their centrally peaked surface brightness profiles. 


A shortcoming that Hydrangea simulations and most cosmological hydrodynamical simulations have is their inability to properly simulate cool-core clusters due to the large entropy values in the cluster cores, which results in flat entropy profiles. An attempt to solve this issue in EAGLE-like simulations has been done by e.g. \citet{2022arXiv221009978A}, however, the issue with the flat entropy profile has not yet been solved completely. Therefore, for the purpose of calculating the background emission, we replace the galaxy clusters in the Hydrangea simulations by observational profiles taken from the sample of cool-core clusters from \citet{2006ApJ...640..691V}. We select $6$ relaxed bright cool-core galaxy clusters (see Table \ref{tab:properties_Vikhlinin_clusters}), which span a range of redshifts, masses, and temperatures. Although our method can in principle be used with any extended source, these galaxy clusters should be among the most promising background sources for our study. \citet{2006ApJ...640..691V} provide the best-fitting parameters for the density and temperature profiles for these galaxy clusters as observed by the Chandra observatory, which were used to calculate the SB profiles as well as SEDs using the galaxy cluster model in SPEX described in the following section (Sec.\,\ref{sec:cluster_model}).


\begin{table}
	\centering       
	\caption{Redshift $z_{\rm cluster}$ and $r_{500}$ of the clusters used in our study taken from \citet{2006ApJ...640..691V}.  $N_{\rm H, TOT}$ is the total hydrogen column density in the direction of the galaxy cluster (see Sec.\,\ref{Sec:Galaxy_absorption} for more details).  }
	\begin{tabular}{|l|c|c|c|c||c}
		\hline
		Cluster & $z_{\rm cluster}$ & $r_{500}$ [kpc] & $N^{\rm neutral}_{\rm H, TOT}$ [cm$^{-2}$] & $100$\,kpc at $z_{\rm cluster}$ \\ \hline
		A2390 & $0.2302$ & $1416$ $\pm$ $48$ &  $8.38 \times 10^{20}$  & $0.45$' \\ \hline
		A383 & $0.1883$ & $944$ $\pm$ $32$ & $3.88 \times 10^{20}$ & $0.53$'  \\ \hline
		A1413 & $0.1429$ & $1299$ $\pm$ $43$ & $1.97 \times 10^{20}$ & $0.66$' \\ \hline
		A2029 & $0.0779$ & $1362$ $\pm$ $43$ &  $3.70 \times 10^{20}$ & $1.13$' \\ \hline
		A1795 & $0.0622$ & $1235$ $\pm$ $36$ &  $1.24 \times 10^{20}$ & $1.39$' \\ \hline
		A262 & $0.0162$ & $650$ $\pm$ $21$ &  $7.15 \times 10^{20}$ & $5.05$' \\ \hline
	\end{tabular}
	\label{tab:properties_Vikhlinin_clusters}
\end{table}

\subsubsection{Galaxy cluster model in SPEX}
\label{sec:cluster_model}

To simulate the spectral energy distribution of the selected background galaxy clusters (used for an approximation of the backlight source), together with their SB profiles (used for weighting the optical depth profiles from the simulations), we use the \emph{cluster} model in SPEX. 

SPEX, the SPEctral X-ray and extreme ultraviolet software package \citep{kaastra1996_spex, kaastra2018_spex, kaastra_j_s_2020_4384188}, is a software package with its own atomic database SPEXACT (The SPEX Atomic Code \& Tables) which includes around $4.2 \times 10^6$ lines from $30$ different chemical elements (H to Zn). It is used for the modelling and analysis of high-resolution X-ray spectra. In this paper we use SPEX version $3.07.02$\footnote{For the most recent version see \url{https://spex-xray.github.io/spex-help/changelog.html}.}. Unless stated otherwise, we use the \citet{2017A&A...601A..85U} ionisation balance and the protosolar abundances by \citet{2009LanB...4B..712L}.


The \emph{cluster} model, which has been newly implemented in SPEX, calculates the spectrum and radial profiles for a spherically symmetric approximation of a cluster of galaxies. It takes as input parametrised 3D radial profiles of the gas density, temperature, metal abundance, and turbulent velocities. Given these parametrised profiles, the emission in multiple 3D shells, each approximated as a single temperature model in collisional ionisation equilibrium (CIE), is computed and projected onto the sky. The cluster model can then be run in different modes, where the output is either the SED within a user-defined projected spatial region, or a radial SB profile in a user-requested energy band. 

More specifically, the assumed shapes of the underlying 3D profiles used as input to the cluster model follow:
\begin{itemize}[leftmargin=5pt]
	\item a two $\beta$ model for the density distribution (with the additional possibility to introduce a jump or break in the slope at a given radius),
	\item the same functional form for the temperature profile as that proposed in \citet{2006ApJ...640..691V}, again with a possible additional jump (although this latter capability was not used in this work),
	\item the abundance profile functional form defined in \citet{2017A&A...603A..80M}.
\end{itemize}
For more details about this model we refer to Štofanová et al. (in preparation) as well as the SPEX manual page\footnote{For now the details of the \emph{cluster} model can be found in the SPEX manual \url{https://var.sron.nl/SPEX-doc/spex-help-pages/models/clus.html}.}.

We can therefore  directly use the fitted parameters already reported in \citet{2006ApJ...640..691V} to describe the 3D temperature profiles of clusters in our sample. For the density profile, on the other hand, the functional form defined in SPEX is different from that adopted in \citet{2006ApJ...640..691V}. Thus, for each target, we attempted to approximate the 3D density profile reported in \citet{2006ApJ...640..691V} with the new functional form available in the \emph{cluster} model. For all six clusters, we were able to match the two functional forms, with differences of at most $15$\% for all radii up to $2\times r_{500}$ . We additionally assume that the metallicity of each cluster follows the average abundance profile of the sample investigated in \citet{2017A&A...603A..80M}. The turbulent velocity is assumed to be $100$\,km/s, which is the current default in the \emph{cluster} model.


\subsubsection{Model for the Milky Way absorption}
\label{Sec:Galaxy_absorption}

In order to realistically simulate the observation of gas in filaments, we need to take into account the absorption by our own Galaxy. The interstellar medium of the Milky Way is a multiphase gas that can not be simply modelled as a neutral gas. Therefore we adopt a model from \citet{2018MNRAS.474..696G}, which assumes the ISM to have three components: (a) a neutral/cold component at temperature $10^4$\,K ($9 \times 10^{-4}$\,keV), (b) a warm component at temperature $10^{4.7}$\,K ($4.3  \times 10^{-3}$\,keV), and lastly (c) a hot component at temperature $10^{6.3}$\,K ($0.17$\,keV). We simulate all of these components with the $hot$ model in SPEX. However, we set our neutral component to temperature $1  \times 10^{-6}$\,keV, which is lower than in \citet{2018MNRAS.474..696G}. The reason is that in SPEX versions $3.06.00$, $3.06.01$ and higher, the database includes charge exchange processes which affect the ionisation at low temperatures, and in such cases the plasma would not be completely neutral. 

The total hydrogen column densities that we use for the neutral component of the ISM, $N^{\rm neutral}_{\rm H, TOT}$, were taken from \citet{2013MNRAS.431..394W}\footnote{We used the online tool created by the SWIFT team \url{https://www.swift.ac.uk/analysis/nhtot/}.} (see Table \ref{tab:properties_Vikhlinin_clusters}). For all galaxy clusters in this paper we assume that the warm and hot components of the ISM have a total hydrogen column density $N^{\rm warm}_{\rm H, TOT} = 0.5 \times 10^{20}$\,cm$^{-2}$ and $N^{\rm hot}_{\rm H, TOT} = 0.2 \times 10^{20}$\,cm$^{-2}$, respectively. These hydrogen column densities were taken from Fig. 4 of \citet{2018MNRAS.474..696G}, which shows that most of the extragalactic sources have quite similar hydrogen column densities.



\begin{figure*}
	\centering
	\includegraphics[width=0.9\textwidth]{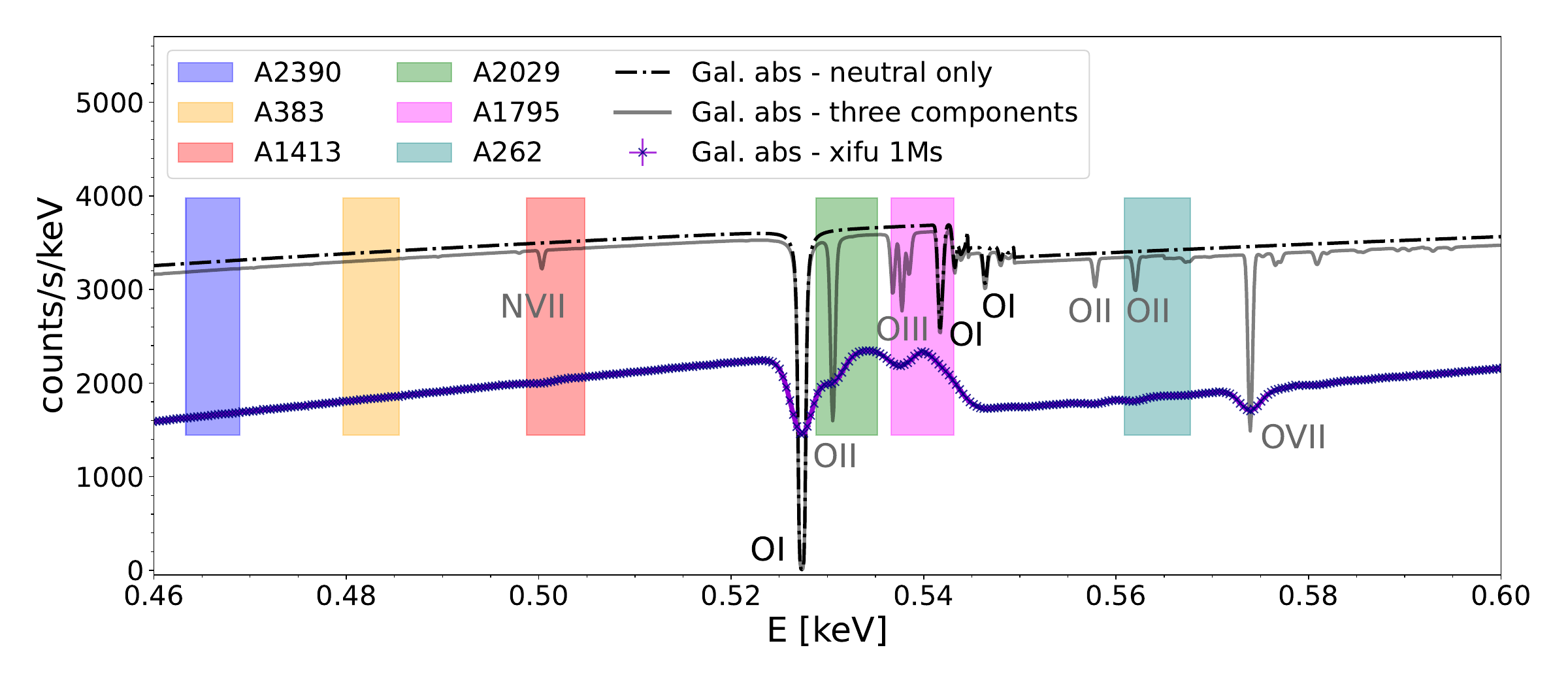}
	\caption{Absorption spectrum of the Galactic ISM for a completely neutral gas (black dash-dotted line) and for the three-component model of ISM (neutral, warm, and hot phases, solid grey line). In the energy range we are interested in ($0.46$--$0.60$\,keV), the neutral component contributes to the spectrum with \ion{O}{I} lines, while the three-component model adds \ion{O}{II}, \ion{O}{III}, \ion{O}{VII} and \ion{N}{VII} lines to the spectrum. The violet solid line with dark blue points (with lower normalization) shows the three-component ISM model folded through the Athena X-IFU response. To demonstrate the effect of the neutral and three-component Galactic absorption model, we simulate the background source of light in this figure as a power law with an arbitrary normalisation and exposure time. The coloured shaded rectangles represent the minimum and maximum energy ranges ($v_{\rm TOT} \sim \pm 1000$\,km/s) of the strongest absorption line profiles from our sample (CE-$25$ and CE-$29$ $z+$) for different galaxy cluster redshifts given in Table \ref{tab:properties_Vikhlinin_clusters}. This figure demonstrates that the ISM lines which might compromise the detection of \ion{O}{VII} cosmic web filaments are \ion{N}{VII}, \ion{O}{I}, \ion{O}{II}, \ion{O}{III}, and \ion{O}{VII}.}
	
	
	\label{Fig:CE_25_all_abell_clusters}
\end{figure*}

Fig.\,\ref{Fig:CE_25_all_abell_clusters} shows the effects of a neutral-only and three-component model of Galactic ISM. For illustration purposes, we plot the effects of absorption on a power-law shaped spectrum; however, for our final analysis the same absorption model was folded with the more complex SED of each cluster. The absorption profile regions represented by shaded rectangles (determined by the total velocity of the strongest absorbers - CE-$25$ and CE-$29$ $z+$) are overplotted for the different cluster redshifts taken from Table \ref{tab:properties_Vikhlinin_clusters}. As we can see, the galaxy cluster redshift and the velocity difference between the galaxy cluster and the absorbing gas in filaments are crucial factors for the \ion{O}{VII} detectability, together with the spectral resolution of the detector. Continuous improvements in determining the column densities for the Galactic ISM are needed to distinguish between the ISM and the \ion{O}{VII} absorption lines. 


\subsubsection{Athena X-ray Integral Field Unit (X-IFU)}
\label{Sec:instruments}

Unlike the grating spectroscopic X-ray instruments that were used to search for the WHIM until now, non-dispersive micro-calorimeters uniquely allow high resolution X-ray spectroscopy of extended sources. 


The Athena X-ray observatory (Advanced Telescope for High Energy Astrophysics) is an L(large)-class mission selected by the European Space Agency (ESA) which plans to study the hot and energetic parts of the Universe, including baryons in the cosmic web that are the focus of this paper. On board it will carry the Wide Field Imager (WFI, \citealp{2014SPIE.9144E..2JM, 2018SPIE10699E..1FM}) and the X-ray Integral Field Unit (X-IFU, \citealp{2018SPIE10699E..1GB}), which is a micro-calorimeter based on a large format array of super-conducting molybdenum-gold Transition Edge Sensors (TESs). The X-IFU will operate in the energy band $0.2$--$12$\,keV and it is expected to have an unprecedented combination of a large effective area of $5900$\,cm$^2$ at $0.5$\,keV (and an even larger effective area of $1.4$\,m$^2$ at $1$\,keV) and a large field of view (FoV) of $5$ arcminutes combined with an angular resolution of $5$ arcseconds, and a spectral resolution of $2.5$\,eV at $0.5$\,keV. For the simulation of data with X-IFU we used the $2018$ response files that can be downloaded here \url{http://x-ifu.irap.omp.eu/resources/for-the-community} \citep{2022arXiv220814562B}. 




\subsubsection{Summary of the methods and the fitting procedure}
\label{sec:summary:obs:methods}

In order to study the detectability of the \ion{O}{VII} and \ion{O}{VIII} absorption lines, we first calculate the surface brightness weighted optical depth profiles of three Hydrangea absorbers: CE-$7$, CE-$25$, and CE-$29$ $z+$. The CE-$25$, and CE-$29$ $z+$ absorbers are the strongest absorbers out of our sample, and additionally to these we also select CE-$7$ in order to simulate less prominent absorptions. For these absorbers we have a full sample of $276$ LoSs which we combine into one final absorption profile by weighting every LoS contribution by the surface brightness profile (in the bolometric energy range) of a given background galaxy cluster. These LoSs were drawn from the circular area with radius $100$\,kpc from the cluster core. We calculate the SB profiles using the \emph{cluster} model, which gives us the bolometric energy flux for every projected shell (the user can specify the energy range in which the flux is calculated). The optical depth in \ion{O}{VII}, and \ion{O}{VIII}, is given by
\begin{equation}
	\tau_{\ion{O}{VII}, \ion{O}{VIII}} = \frac{\sum_{m = 0}^{276} \, \textrm{SB}{(R)} \times \tau^m_{\ion{O}{VII}, \ion{O}{VIII}}{(R)}  }{ \sum_{m = 0}^{276} \, \textrm{SB}{(R)} } \;,
	\label{eq:SB_weighting}
\end{equation}
where index $m$ represents the individual LoSs, and $R$ is the projection radius. We convolve the final weighted Hydrangea absorption profile with the galaxy cluster SEDs (also calculated for the projected radius of $100$\,kpc). We take into account the three-component absorption model for the ISM of the Milky Way and redshift the \ion{O}{VII} and \ion{O}{VIII} Hydrangea absorption lines with respect to the redshift of the background galaxy clusters. These spectra are then folded through the response files of the Athena X-IFU to simulate the observations with different exposure times ($100$\,ks, $250$\,ks, and $500$\,ks) for different absorbers. 

We fit the simulated spectra in SPEX in a narrow energy band around the absorption line, which is typically $0.4-0.6$\,keV. For simplicity, we fit the background continuum with a power law. We can afford this assumption, since in this narrow energy band the clusters do not seem to have many emission lines. If the spectrum of a specific cluster (as e.g. A$262$ or A$383$) has strong emission lines in the energy band $0.4-0.6$\,keV, we exclude the energy intervals corresponding to these emission lines during the fitting procedure. Since the shape of the Hydrangea absorption profiles is unresolved by X-IFU ($2.5$\,eV), we fit the absorption lines from Hydrangea with a gaussian line. If the Hydrangea absorption lines are not sufficiently separated from Galactic lines of the local ISM and the peaks of the lines cannot be clearly resolved, we discard these simulated observations from our sample. If the Hydrangea absorption line is sufficiently separated from the ISM lines but within $2-6$\,eV from each other, we leave the abundance of the element in that particular ISM component free during the fitting procedure (e.g. as in case of Abell $2029$ and Abell $262$). In such cases we do not only calculate the errors on the normalisation of the absorption line, but also the errors on the fitted abundance of that particular element. The errors are estimated by using the Levenberg-Marquardt algorithm and represent the standard deviation of $1\sigma$. We use C-statistics \citep{1979ApJ...228..939C}, which can be briefly summarised as the maximum likelihood estimation in the limit of Poissonian statistics. SPEX uses modified C-statistics based on \citet{1984NIMPR.221..437B}, which is described in detail in \citet{2017A&A...605A..51K}. The fit is considered good, if the C-statistics value of the fit is within/close to the range of the expected C-statistics value and its errors.


\subsection{Results: Simulations with the Athena X-IFU}
\label{Sec:XIFU_results}

\subsubsection{Significance of the \ion{O}{VII} detection}
\label{sec:OVII}

\begin{figure*}
\centering
\begin{subfigure}[c]{\textwidth}
	\centering
    \includegraphics[width=0.3\textwidth]{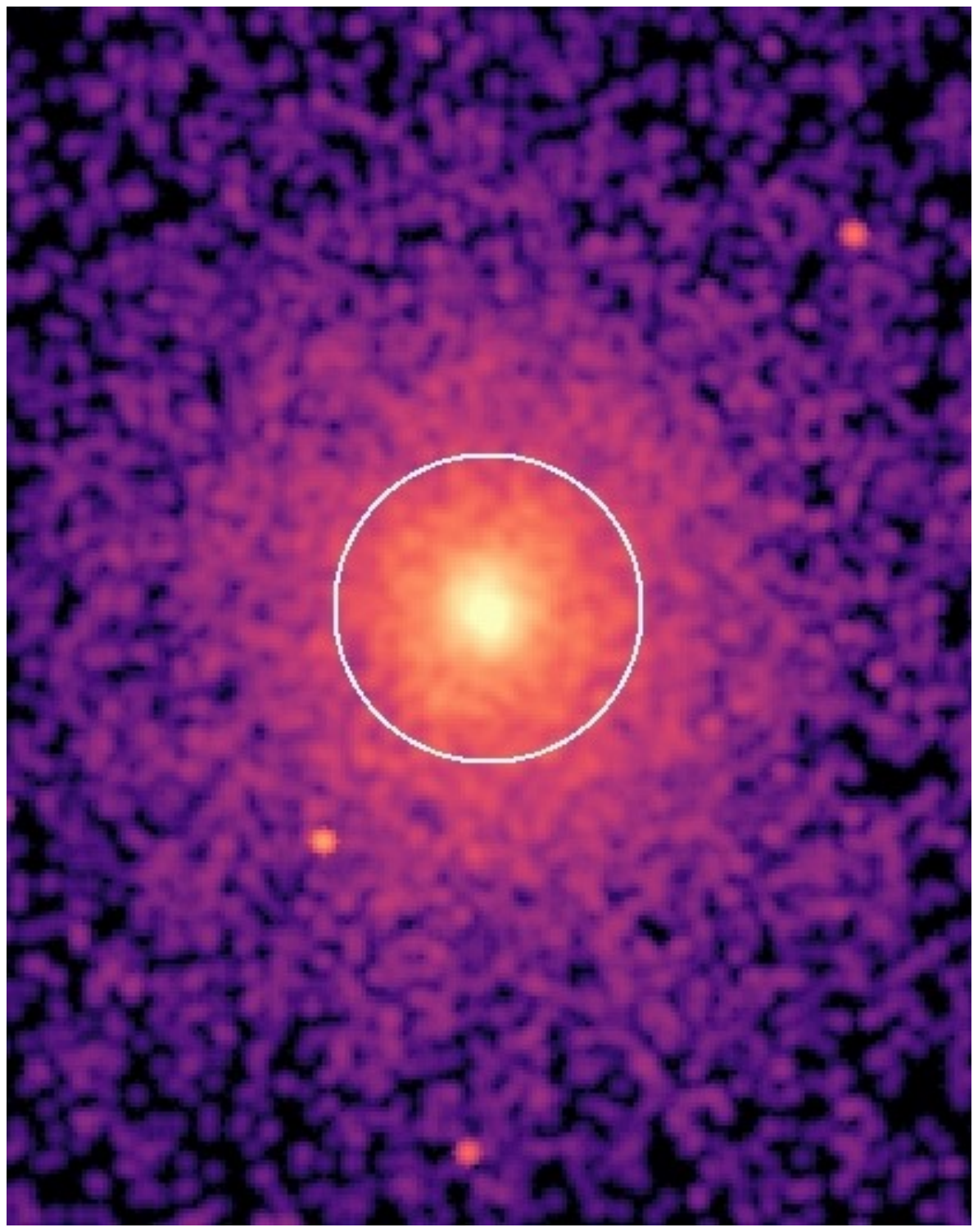}
    \includegraphics[width=0.58\textwidth]{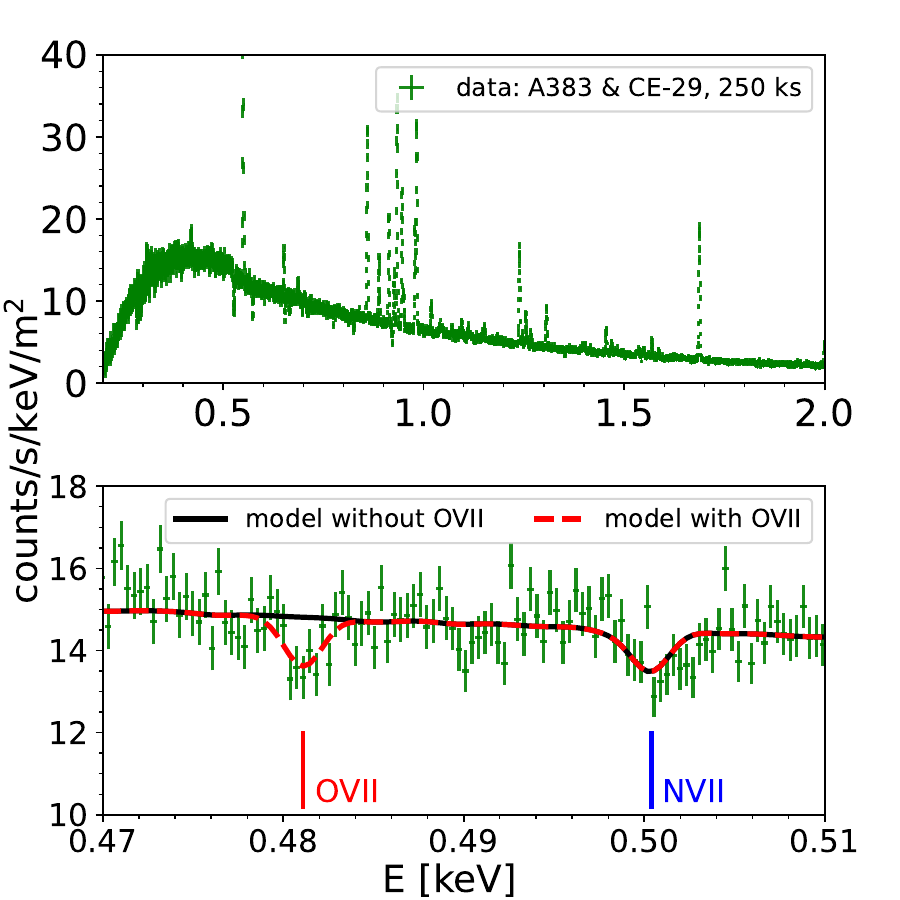}
\end{subfigure}%

\begin{minipage}[c]{\textwidth}
    \centering
	\includegraphics[width=0.3\textwidth]{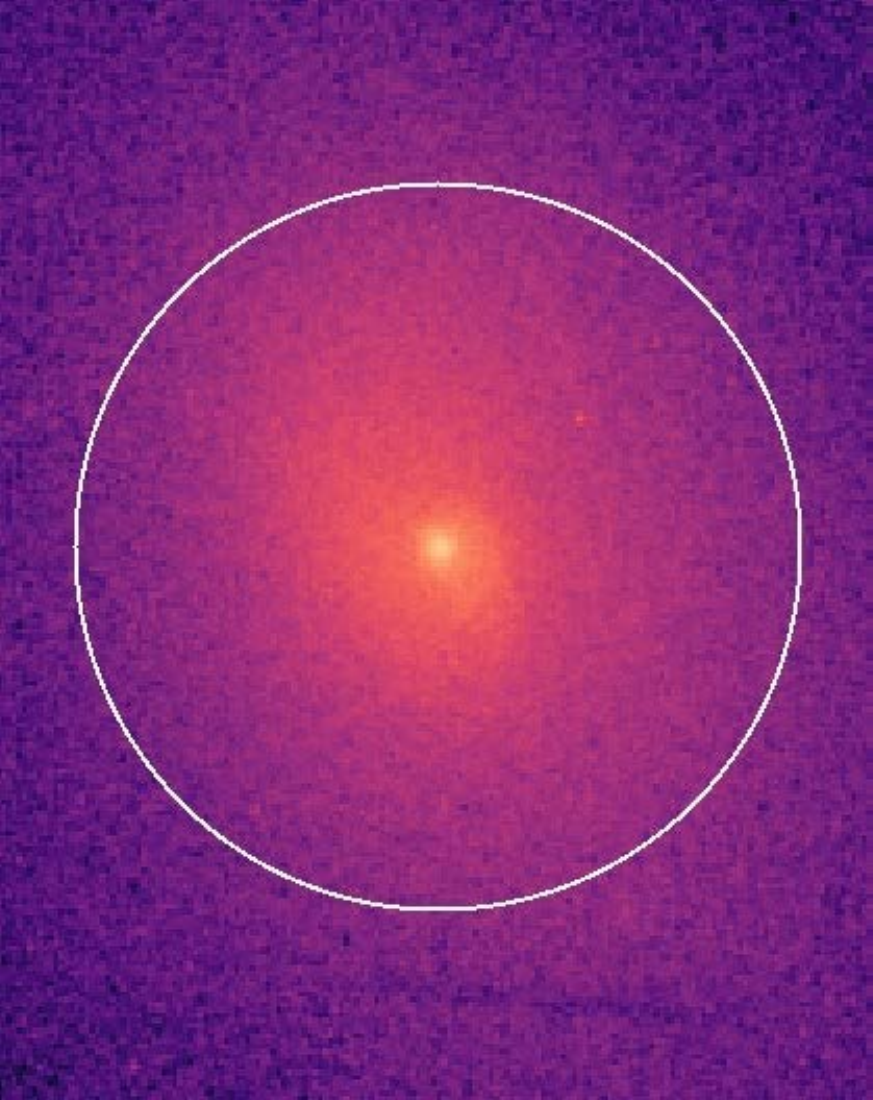}
	\includegraphics[width=0.58\textwidth]{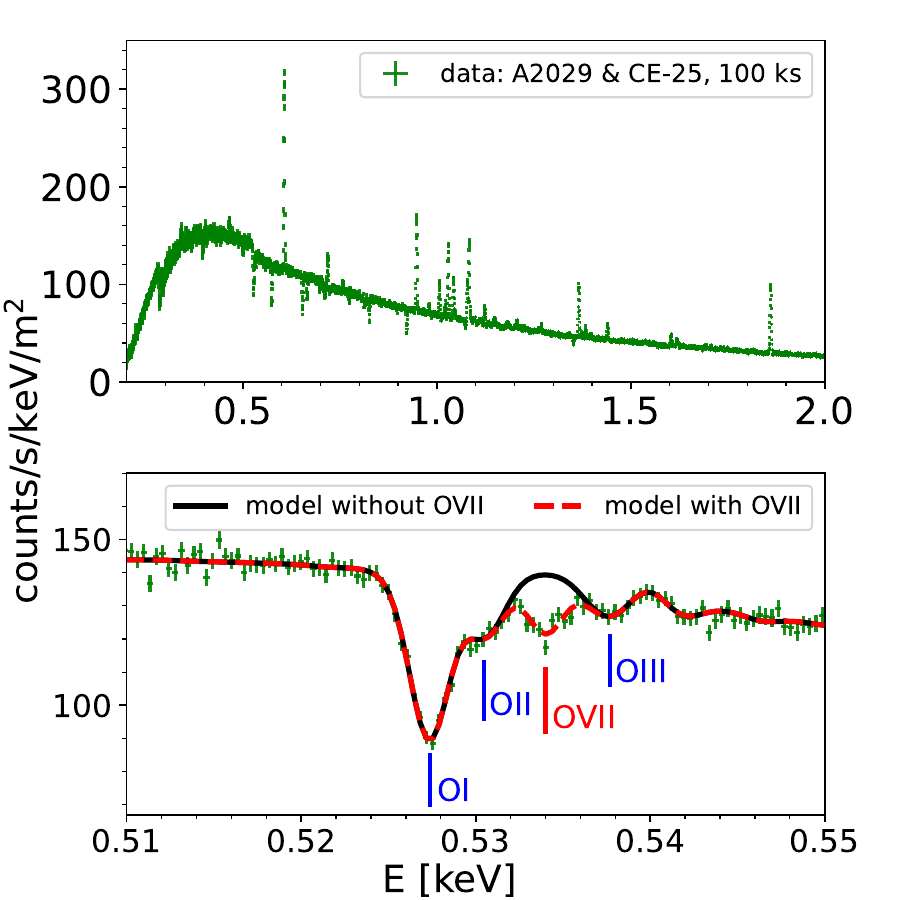}
\end{minipage}
\caption{Broad band simulated spectrum for Athena X-IFU in the energy range $0.2-2$\,keV and a zoom-in to $0.47-0.51$\,keV of the CE-$29$ $z+$ Hydrangea absorber and A$383$ (top two right panels, $250$\,ks exposure time, $z=0.1883$) and the CE-$25$ and A$2029$ (bottom two right panels, $100$\,ks exposure time, $z=0.0779$), where the zoom-in for CE-$25$ and A$2029$ shows the energy band $0.51-0.55$\,keV. The black solid line represents the spectral model (folded through the response of X-IFU) together with the three-component absorption model of the Milky Way ISM as described in Sec.\,\ref{Sec:Galaxy_absorption}. The absorption lines of the local ISM are indicated in all panels with blue vertical lines. The red dashed line represents the same spectral model as the black solid line but with an addition of the Hydrangea \ion{O}{VII} absorption line (solid red vertical lines) according to the best-fitting parameters described in Table \ref{Tab:fitting_results}. Left panels show the Chandra images of A$383$ (top left panel) and A$2029$ (bottom left panel). The white circle has a radius of $100$\,kpc as used in all calculations in this paper. The errors shown in the spectra represent the standard deviation $1\sigma$.}	
	\label{Fig:Simulated_spectra}
\end{figure*}

In Fig.\,\ref{Fig:Simulated_spectra} we show X-IFU simulated spectra of the \ion{O}{VII} Hydrangea absorption profiles for the CE-$29$ Hydrangea absorber and A$383$ background galaxy cluster at redshift $z=0.1883$ (also referred to as CE-$29$ \& A$383$) for an exposure time of $250$\,ks. We also show CE-$25$ \& A$2029$ at redshift $z=0.0779$ and $100$\,ks exposure time. These two cases represent the 'best' and the 'worst' scenarios for \ion{O}{VII} observations that are still possible with X-IFU. The left panels in Fig.\,\ref{Fig:Simulated_spectra} are Chandra images of A$383$ (top left panel, Chandra observations ID $2321$\footnote{A$383$ Chandra observations with ID $2321$ can be downloaded from \url{https://doi.org/10.25574/02321}.}) and A$2029$ (bottom left panel, Chandra observations ID $4977$\footnote{A$2029$ Chandra observations with ID $4977$ can be downloaded from \url{https://doi.org/10.25574/04977}.}).

The 'best' possible scenario for \ion{O}{VII} observations is when the Hydrangea absorption line is sufficiently separated from any of the ISM lines. We show as an example X-IFU simulated spectra of the CE-$29$ absorber projected in front of the galaxy cluster A$383$. For all three chosen Hydrangea absorbers (CE-$7$, CE-$25$, and CE-$29$ $z+$) and for A$383$ and A$2390$, the \ion{O}{VII} Hydrangea absorption line is sufficiently separated from the ISM lines and can potentially be observed with X-IFU. 


The 'worst' possible scenario for \ion{O}{VII} observations that is still feasible with X-IFU is when the Hydrangea absorption line is blended with ISM lines, but its peak is sufficiently separated from the ISM lines and can be resolved with X-IFU. This is the case for, for example, CE-$25$ \& A$2029$, where the Hydrangea \ion{O}{VII} absorption line is close to the complex of oxygen lines from the Galactic ISM as shown in Fig.\,\ref{Fig:Simulated_spectra}. These lines are the \ion{O}{I} line from the neutral component, and the \ion{O}{II} and \ion{O}{III} lines from the warm component. In this case we fit the \ion{O}{VII} absorption line while allowing the abundance of oxygen in the neutral as well as the warm component of the ISM to be free during fitting. For two other Hydrangea absorbers, CE-$7$ and CE-$29$, the velocity difference between the galaxy cluster A$2029$ and the gas in the filaments is not 'favourable' and the Hydrangea absorption lines are not sufficiently separated from the \ion{O}{II} ISM line. 

As already seen from Fig.\,\ref{Fig:CE_25_all_abell_clusters}, for A$1795$ the \ion{O}{VII} line lies close to the \ion{O}{I} or \ion{O}{III} ISM lines, as well as close to the oxygen absorption edge. With the X-IFU simulations we confirm that none of the three chosen examples of Hydrangea absorbers are sufficiently separated from the ISM lines and we discard these cases from our sample.  

For the case of A$1413$ one needs to be careful about the \ion{N}{VII} line from the hot component of the ISM at $0.5$\,keV. Two of the three absorbers considered here (CE-$7$ and CE-$29$) are too close to this line to be resolved with the X-IFU.
The column density of the hot component is not high enough for the \ion{N}{VII} line to be significant in the Athena spectrum, which makes fitting this line even more difficult. Therefore, while fitting the \ion{O}{VII} line of the third absorber (CE-$25$), we fix the nitrogen abundance to the initial input value (which is equal to unity for the \citealp{2009LanB...4B..712L} proto-solar abundances) during the fitting procedure. However, the column density of the hot phase ISM is not well known and in reality the \ion{N}{VII} line may be more significant in the observed spectra of the galaxy cluster.

Table \ref{Tab:fitting_results} contains the results of fitting for three Hydrangea absorbers and all galaxy cluster background sources mentioned in Table \ref{tab:properties_Vikhlinin_clusters}. From this table we see that even in the case of the less prominent absorbers (here represented by CE-$7$), for which the flux in \ion{O}{VII} does not decrease by more than $40$\% (see Fig.\,\ref{Fig:tau_vs_vtot_vs_energy}), \ion{O}{VII} can be detected with X-IFU with at least $5\sigma$ significance. This is possible with an exposure time of  $385$\,ks for Abell $2390$ and $195$\,ks for Abell $383$ (see Table \ref{Tab:fitting_results_exposure}).  Unfortunately, for closer and brighter galaxy clusters in our sample, the CE-$7$ absorption line of \ion{O}{VII} cannot be resolved from  the \ion{N}{VII}, \ion{O}{II} or \ion{O}{III} ISM lines. 


More prominent absorbers, which show depletion of the flux in \ion{O}{VII} of more than $70$\%, and which have a significant velocity difference between the gas in the filaments and the galaxy cluster core (in our studies this difference is $v_{\rm TOT} \sim \pm 1000$\,km/s), can almost all (besides CE-$25$ \& A$2390$ and CE-$29$ \& A$1413$) be detected with more than $5\sigma$ significance with exposure time $\leq 250$\,ks (for high redshifts, such as Abell $383$) or even $100$\,ks for the lower-redshift galaxy clusters, such as Abell $2029$ or Abell $262$. The CE-$25$ absorber in our study is always resolved from the ISM lines while CE-$29$ can potentially be observed only for high-redshift clusters such as A$2390$ and A$383$. 

\begin{table*}
	\caption{X-IFU simulations for the CE-$7$, CE-$25$, and CE-$29$ $z+$ Hydrangea \ion{O}{VII} absorbers and the A$2390$, A$383$, A$1413$, A$2029$, A$1795$, and A$262$ background galaxy clusters. $t_{\rm exp}$ is the exposure time and the normalisation of the Gaussian line is given in units of $\times 10^{49}$\,ph/s. We also provide the values of the C-statistics, the significance of the detection $\sigma$, as well as the value of the expected C-statistics (see Sec.\,\ref{sec:summary:obs:methods} and Sec.\,\ref{sec:OVII} for more details). The radius of interest was assumed to be $100$\,kpc and the final absorption profiles were weighted by the surface brightness of the respective galaxy cluster. The errors represent the standard deviation $1 \sigma$.}
	\begin{tabular}{|l|l|c|c|c|}
		\hline
		&  & \textbf{CE-7} & \textbf{CE-25} & \textbf{CE-29} \\ \hline
		\textbf{Abell 2390} 	& $t_{\rm exp}$ [ks] & 500 & 250 & 250 \\		
		& normalisation [$\times 10^{49}$\,ph/s] & $-8.6 \pm 1.5 $ &  $-8.5 \pm 2.0 $ &  $-15.9 \pm 2.1 $ \\ 
		
		\textbf{} & C-statistics & $829$ & $799$ & $757$ \\ 
		\textbf{} & expected C-statistics &  $791 \pm 40 $ &  $791 \pm 40 $ &  $791 \pm 40 $ \\ 
		\textbf{} & significance $\sigma$ & \textbf{5.7} & \textbf{4.3} & \textbf{7.6} \\ \hline 		
		
		\textbf{Abell 383} 	& $t_{\rm exp}$ [ks] & 500 & 250 & 250 \\		
		& normalisation [$\times 10^{49}$\,ph/s] & $-4.8 \pm 0.6 $ & $-4.5 \pm 0.9$ & $-6.4 \pm 0.9$  \\ 
		\textbf{} & C-statistics & $653$ & $649$ & $669$ \\ 
		\textbf{} & expected C-statistics & $588 \pm 34$ & $588 \pm 34$ & $588 \pm 34$ \\ 
		\textbf{} & significance $\sigma$ & \textbf{8.0} & \textbf{5.0} & \textbf{8.0} \\ \hline

		\textbf{Abell 1413} & $t_{\rm exp}$ [ks] &  & 250 &   \\ 
		& normalisation [$\times 10^{49}$\,ph/s] & & $-4.1 \pm 0.5$  &  \\ 
		\textbf{} & C-statistics & blended with & $415$ & blended with \\ 
		\textbf{} & expected C-statistics &  \ion{N}{VII} & $419 \pm 29$ &  \ion{N}{VII} \\ 
		\textbf{} & significance $\sigma$ & & \textbf{8.2} &   \\ \hline 
		
		\textbf{Abell 2029} & $t_{\rm exp}$ [ks] &  & 100 &  \\ 
		& normalisation [$\times 10^{49}$\,ph/s] & & $-9.1 \pm 0.5$ &  \\ 
		\textbf{} & abundance O (neutral) &  & $1.01 \pm 0.02$ &  \\ 
		\textbf{} & abundance O (warm) & blended with & $1.06 \pm 0.05$ & blended with \\  
		\textbf{} & C-statistics & \ion{O}{II} & $150$ & \ion{O}{II} \\ 
		\textbf{} & expected C-statistics &  & $172 \pm 19$ &  \\ 
		\textbf{} & significance $\sigma$ &  & \textbf{18.2} &  \\ \hline 
		
		\textbf{Abell 1795} & \textbf{} & blended with \ion{O}{III} & blended with \ion{O}{I} & blended with \ion{O}{III} \\ \hline 
		
		\textbf{Abell 262} & $t_{\rm exp}$ [ks] &  & 100 &  \\ 
		& normalisation [$\times 10^{49}$\,ph/s] &  & $-0.23 \pm 0.02$ &  \\ 
		\textbf{} & abundance O (hot) & blended with & $1.3 \pm 0.2$ & blended with \\ 
		\textbf{} & C-statistics & \ion{O}{II} & $408$ & \ion{O}{II} \\ 
		\textbf{} & expected C-statistics &  & $430 \pm 29$ &  \\ 
		\textbf{} & significance $\sigma$ &  & \textbf{11.5} &  \\ \hline
	\end{tabular}
	\label{Tab:fitting_results}
\end{table*}



The exposure times needed to get a $5\sigma$ detection in \ion{O}{VII} with Athena X-IFU for CE-$7$, CE-$25$, and CE-$29$ $z+$ absorbers are summarized in Table \ref{Tab:fitting_results_exposure}.

\begin{table}
	\centering       
	\caption{Exposure time $t_{\rm exp}$ in ks for X-IFU ($2018$ baseline response matrices) simulations for CE-$7$, CE-$25$, and CE-$29$ $z+$ Hydrangea \ion{O}{VII} absorbers and A$2390$, A$383$, A$1413$, A$2029$, A$1795$, and A$262$ background galaxy clusters for a detection of $5\sigma$. The radius of interest was assumed to be $100$\,kpc and the final absorption profiles were weighted by the surface brightness profile of the respective galaxy cluster. }
	\begin{tabular}{|c|c|c|c|}
		\hline
		$t_{\rm exp}$ [ks] & \textbf{CE-7} & \textbf{CE-25} & \textbf{CE-29} \\ \hline
		\textbf{Abell 2390} &  385 & 338 & 108 \\		
		
		\textbf{Abell 383} 	&  195 & 250 & 98 \\		
		
		\textbf{Abell 1413} & -- & 93 & -- \\ 
		
		\textbf{Abell 2029} & -- & 8 & -- \\ 
		
		\textbf{Abell 1795} & -- & -- & -- \\
		
		\textbf{Abell 262} & -- & 19 & -- \\  \hline
	\end{tabular}
	\label{Tab:fitting_results_exposure}
\end{table}

\subsubsection{Significance of the \ion{O}{VIII} detection}
\label{sec:OVIII} 

The detection of the cosmic web filaments against galaxy cluster background sources in \ion{O}{VIII} is significantly more difficult than for \ion{O}{VII}. Firstly, the depletion of flux in \ion{O}{VIII} is comparable to the less prominent \ion{O}{VII} absorbers, which as we have shown in Table \ref{Tab:fitting_results}, require more exposure time for a significant detection (more than $200$\,ks or $300$\,ks).  For most of these absorbers the flux in  \ion{O}{VIII} decreased by $10-20$\% relative to the continuum (see Fig.\,\ref{Fig:tau_vs_vtot_vs_energy}). For some absorbers, as e.g. CE-$3$, CE-$8$, the line flux decreased by $35$\%. Secondly, due to high temperatures and densities in the intracluster medium (ICM), the galaxy cluster spectra contain a strong \ion{O}{VIII} emission line. Even though the velocity difference between ICM and the gas in the cosmic web filaments is similar for \ion{O}{VII} and \ion{O}{VIII} absorption (see Fig.\,\ref{Fig:tau_vs_vtot_vs_energy}), the wings of the \ion{O}{VIII} emission line are too bright for the \ion{O}{VIII} absorption line to be sufficiently separated with the X-IFU resolution of $2.5$\,eV. In Sec.\,\ref{Sec:discussion_detectors} we discuss how these results would be affected if we used a detector with an even higher spectral resolution. 


\section{Discussion}
\label{Sec:discussion}


We have shown above that detecting the WHIM in absorption against bright cluster cores is possible (and perhaps even promising) with future, high-resolution, non-dispersive X-ray spectrometers. Below, we discuss in more detail various additional aspects that can affect the detectability of \ion{O}{VII} and \ion{O}{VIII} absorption lines, such as the optimal size of the spectral extraction region, the effect of using background sources with different underlying surface brightness distributions, the impact of the detector's spectral resolution, the impact of the \ion{O}{VII} emission from the cosmic web filaments, and possible blending with emission lines from low-temperature gas in cluster cores.

\subsection{The optimal size of the spectral extraction region}
\label{Sec:discussion_radius}

Until now, we focused our calculations on a circular aperture with a fiducial radius of $100$\,kpc. However, it is not clear if this is the optimal choice for the spectral extraction region. To answer the question of how the selection of the radius of interest affects our results, we did a new set of calculations with \textit{Specwizard} for the  CE-$25$ Hydrangea absorber for two additional radii: $40$\,kpc and $300$\,kpc. As already shown in Table \ref{Tab:fitting_results}, this absorber was detected with Athena for most of the background sources, and therefore we chose to show the results of this subsection for this absorber.  How CE-$25$ looks like on these scales can be seen in Fig.\,\ref{Fig:column_density_maps} (red lines with different line styles).

\begin{figure}
	\centering
	\includegraphics[width=0.5\textwidth]{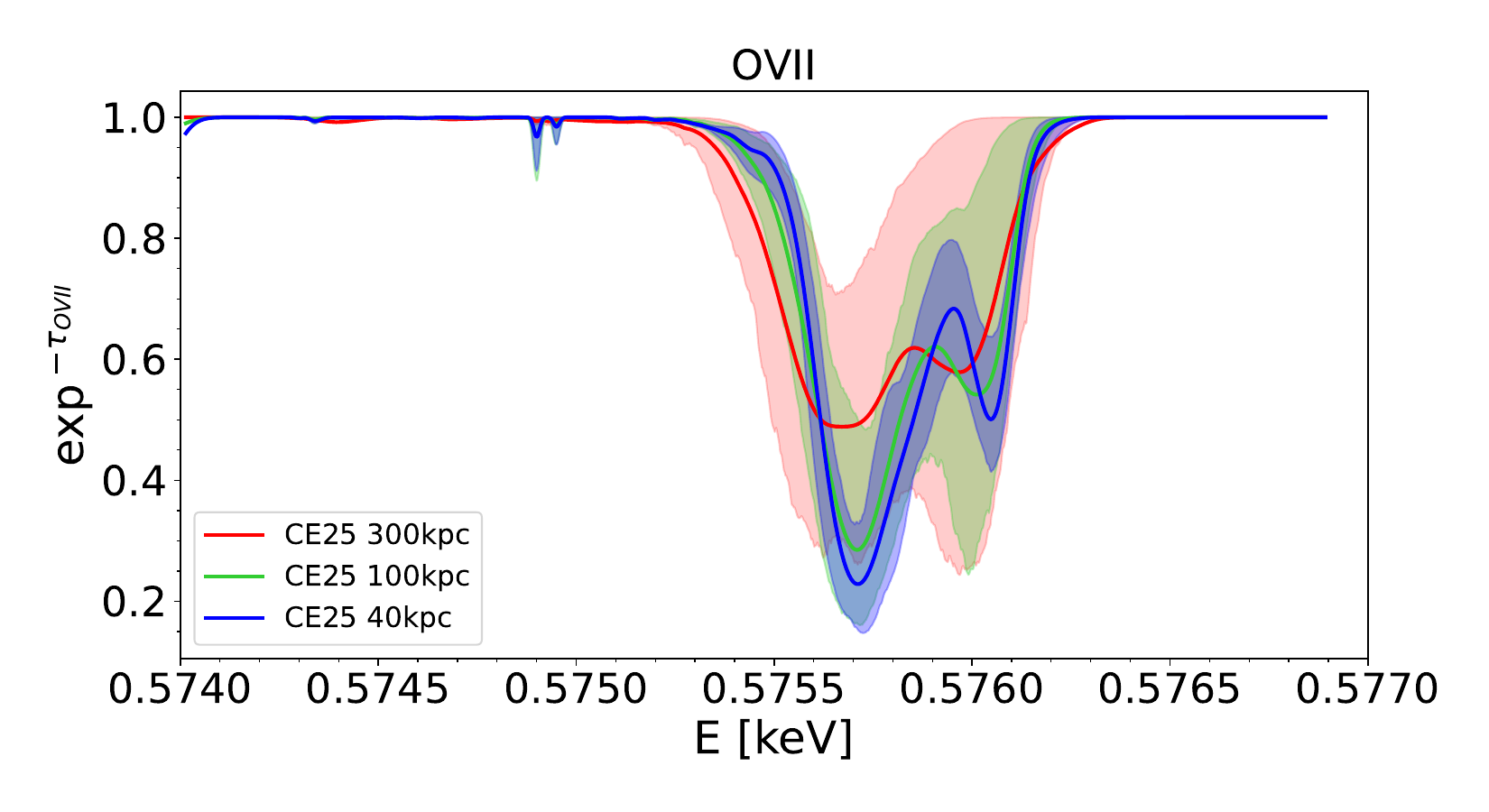}
	\caption{CE-$25$ \ion{O}{VII} absorption profile (equally weighted) at \ion{O}{VII} rest energy for $40$\,kpc (blue solid line), $100$\,kpc (green solid line), and $300$\,kpc (red solid line) spectral extraction regions. In every energy bin, the shaded area represents all the data between the $10$th and $90$th percentiles of the full sample, which consists of $276$ LoSs for each radius. The larger spectral extraction region with $R = 300$\,kpc results in shallower absorption profiles in comparison with $R = 40$\,kpc. }
	\label{Fig:quasar_like_studies_compar}
\end{figure}

The shaded areas in Fig.\,\ref{Fig:quasar_like_studies_compar} show the 10th and 90th percentile scatter between different individual sightlines (from the full sample of $276$ LoSs) covering each radius. Our results indicate that the absorption profiles are deeper for a smaller radius of interest; for larger radii, the scatter among different sightlines increases, and their average absorption profile becomes shallower. The relatively small scatter between individual sightlines probing the central $40$\,kpc radius (shaded blue area in Fig.\,\ref{Fig:quasar_like_studies_compar}) suggests that the projected properties of the CE-$25$ absorber are not very patchy over this spatial scale, while their spatial substructure increases considerably when considering a larger, $300$\,kpc region. Fig.\,\ref{Fig:quasar_like_studies_compar} also provides a comparison to quasar-like studies, where the absorbers would be probed along a single line of sight. In this case, the absorption profile could lie anywhere in the plotted shaded area, whereas the absorption profile for an extended source would follow the solid line.


For calculating the significance of the \ion{O}{VII} detection with X-IFU and quantifying the comparison of different radii of interest, we chose as an example CE-$25$ (SB weighted profile) and A$1413$ as a background source. We again assumed that the absorption profile can be fitted with a gaussian. The results can be found in Table \ref{Tab:fitting_results_radius} for an exposure time of $250$\,ks. 



In conclusion, there is a clear trade-off in choosing the size of the spectral extraction region: the smaller the region, the deeper the absorption profile is likely to be (see Fig.\,\ref{Fig:quasar_like_studies_compar}).  When the region of interest increases, we start to average over a larger variety of sightline properties, which dilutes the average signal. However, larger regions of interest also mean a larger number of counts from the extended background source, and therefore better statistics in determining the continuum level with respect to which the line is measured, increasing the significance of the detection. Our study suggests that, at least out to the radii of interest considered in our study, the latter effect is more important, i.e. the significance of \ion{O}{VII} detection increases with the size of the radius (see Table\, \ref{Tab:fitting_results_radius}). However, we would like to point out that calibration uncertainties (such as gain variation across the detector) may play an additional role when determining the significance of the \ion{O}{VII} detection for various spectral extraction regions.  



\begin{table*}
	\caption{X-IFU simulations for CE-$25$ and the A$1413$ background galaxy cluster for three different spectral extraction regions: $40$\,kpc, $100$\,kpc, and $300$\,kpc. The final absorption profiles were weighted by the surface brightness of A$1413$ and fitted with a gaussian profile. The errors represent the standard deviation of $1 \sigma$. The exposure time for all radii was $250$\,ks. In conclusion, even though the larger radius of interest, $R = 300$\,kpc, results in shallower absorption profiles in comparison with $40$\,kpc, the increase in the number of counts for the larger area improves the significance of the \ion{O}{VII} detection. }
	\begin{tabular}{l|c|c|c}
		\hline
		radius [kpc]& 40 & 100 & 300 \\ \hline
		normalisation [$\times 10^{49}$\,ph/s] & $-1.17 \pm 0.25 $ & $-4.1 \pm 0.5$  & $-9.3 \pm 0.8 $  \\ 
		C-statistics & $482$ & $415$ & $399$ \\ 
		expected C-statistics &  $419 \pm 29$ & $419 \pm 29$ & $419 \pm 29$ \\ 
		significance $\sigma$ & \textbf{4.7} & \textbf{8.2} & \textbf{11.6} \\ \hline 
	\end{tabular}
	\label{Tab:fitting_results_radius}
\end{table*}

\subsection{Impact of weighting by different surface brightness profiles}
\label{Sec:discussion_weight}

In this paper, we demonstrated the method of observing cosmic web filaments against bright cool-core galaxy clusters. We selected a few galaxy clusters which span a range of redshifts, masses, and temperatures. However, one could potentially use any bright cool-core cluster. What would change is the SB profile that is used for weighting of the LoSs that are used for the final Hydrangea absorption profile.
 
To assess the impact of this weighting, we compare the final weighted profiles for two cases: (a) equally weighted profiles (the arithmetic mean), and (b) surface brightness weighted profiles (see Eq.\,\eqref{eq:SB_weighting} in Sec.\,\ref{sec:summary:obs:methods}). The difference between the equally weighted profiles and the SB weighted profiles is mostly $<5$\% and maximally $10$\% for all $16$ absorbers and all galaxy clusters, with the exception of three Hydrangea absorbers which show a difference larger than $10$\%: CE-$1$ $x+$, CE-$22$ and CE-$25$. For these three outliers the difference between the weights depends on the background galaxy cluster, and is the highest for A$383$: $17$\% for CE-$1$ $x+$, and $36$\% for  CE-$22$.

As already seen from the column density maps, the cosmic web filaments in \ion{O}{VII} can be very patchy in the projection on the sky, and they do not have to homogeneously cover the whole spectral extraction region, from which we draw the LoSs for the final absorption profiles. Even though the difference between the arithmetic mean and SB weighting is not as significant, we chose to use the SB weighting for the results obtained in Sec.\,\ref{Sec:XIFU_results}. 



\subsection{Prospects for detectors with a different spectral resolution and effective area}
\label{Sec:discussion_detectors}


In the case of the Athena mission, which is planned for the second half of the $2030$s, one needs to keep in mind that it might still undergo changes in the mission requirements. This could possibly result in the degradation of the effective area or spectral resolution. The change in the effective area would affect mainly the amount of exposure time needed for a detection above $5\sigma$, however, the observations would still be possible and the method and the results presented in this paper would not be affected. The more concerning would be the degradation of the spectral resolution. In the case of \ion{O}{VII}, we tried to simulate data with a spectral resolution of $4$\,eV instead of $2.5$\,eV. For the background sources at a redshift similar to A$2029$, where the Hydrangea absorbers are blended with the \ion{O}{I}, \ion{O}{II}, and \ion{O}{III} ISM lines, such a resolution is no longer enough to sufficiently separate the \ion{O}{VII} Hydrangea line from the ISM lines. We also simulated two other cases from Table \ref{Tab:fitting_results}: (a) CE-$25$ and A$1413$ for a $250$\,ks exposure time, and (b) CE-$25$ and A$262$ for a $100$\,ks exposure time. In the case of A$1413$, the significance of the detection of \ion{O}{VII} decreased from $8.2\sigma$ to $6.3\sigma$. For A$262$ the significance of the detection decreased from $10\sigma$ to $6\sigma$. This means that if the spectral resolution is degraded from $2.5$\,eV to $4$\,eV, the exposure time for a $5\sigma$ detection of the CE-$25$ \ion{O}{VII} absorber would increase from $93$\,ks to $174$\,ks in the case of the background galaxy cluster A$1413$, and from $25$\,ks to $63$\,ks in the case of A$262$. Nevertheless, it is encouraging that such studies would remain feasible.



Another detector that we discuss in this section is the Integral Field Unit (IFU) micro-calorimeter of the Line Emission Mapper (LEM, \citealp{2022arXiv221109827K}). LEM is a mission concept to be submitted to the National Aeronautics and Space Administration (NASA) 2023 Astrophysics Probes call for proposals. It is an X-ray probe that will, among other applications, focus on galaxy formation and evolution by probing the circumgalactic and intergalactic media. It will consist of an X-ray mirror with an effective area of $1600$\,cm$^2$ at $0.5$\,keV, and a cryocooled array of Transition-Edge Sensor (TES) micro-calorimeters, whose technology will be built on Athena X-IFU and the Lynx X-ray observatory concept \citep{2018AAS...23110304V}. 



LEM's biggest advantage considering our scientific interest lies in its spectral resolution of approximately $0.9$\,eV for extended sources. In Fig.\,\ref{Fig:spectrum_A2029_CE25_LEM}, we plot LEM simulated spectrum for CE-$25$ \& A$2029$ \ion{O}{VII} absorption line for a $100$\,ks exposure time and CE-$25$ \& A$2029$ \ion{O}{VIII} absorption line for a $1$\,Ms exposure time. In comparison with X-IFU (see Fig.\,\ref{Fig:Simulated_spectra}), the \ion{O}{I}, \ion{O}{II}, and \ion{O}{III} Galactic absorption lines are completely separated from the \ion{O}{VII} Hydrangea absorption line. We compared these simulated results to the X-IFU results given in Table \ref{Tab:fitting_results}. The significance of the \ion{O}{VII} detection decreased from $18.2 \sigma$ to $15.4 \sigma$, which is caused mainly by the fact that the exposure time was the same for both instruments, even though the effective area is smaller for LEM. In general, the observations with LEM might need more observing time to reach the same significance of detection, however, its unprecedented spectral resolution can increase the number of detections by distinguishing more easily between the absorption from cosmic web filaments and the absorption by the Galactic ISM. In the case of \ion{O}{VIII} (see the bottom panel of Fig.\,\ref{Fig:spectrum_A2029_CE25_LEM}), a spectral resolution comparable to that of LEM can distinguish between galaxy cluster \ion{O}{VIII} emission line and the cosmic web filament \ion{O}{VIII} absorption line. According to our simulations, a detection of $10 \sigma$ is possible in $1$\,Ms. Therefore, a $5 \sigma$ detection would be feasible with LEM for an exposure time of $250$\,ks.


\begin{figure}
	\centering
	\includegraphics[width=0.5\textwidth]{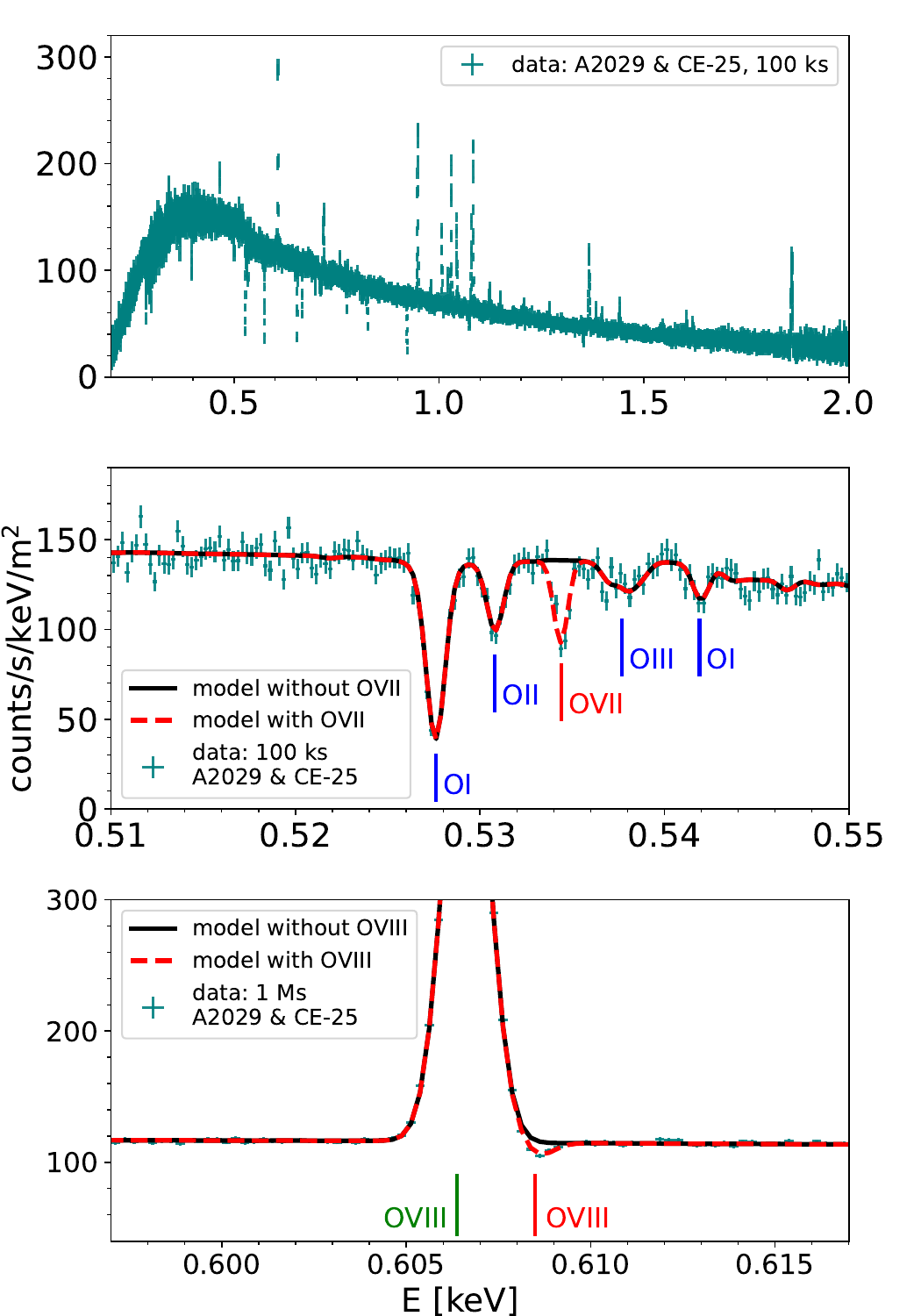}	
	\caption{LEM simulated broad band spectrum in the energy range $0.2-2$\,keV (top panel) and a zoom in to $0.51-0.55$\,keV (middle panel) and $0.6-0.615$\,keV (bottom panel) of the CE-$25$ Hydrangea absorber and backlight cluster A$2029$. The black solid line represents the spectral model (folded through the response of LEM) together with the three-component absorption model of the Milky Way ISM as described in Sec.\,\ref{Sec:Galaxy_absorption}. The absorption lines of the local ISM are indicated in all panels with blue vertical lines. The red dashed line represents the same spectral model as the black solid line but with an addition of the Hydrangea \ion{O}{VII} or \ion{O}{VIII} absorption line (labelled with solid red vertical lines) according to the best-fitting parameters described in Sec.\,\ref{Sec:discussion_detectors}. The green vertical line in the bottom panel shows the \ion{O}{VIII} galaxy cluster emission line.}

	

	\label{Fig:spectrum_A2029_CE25_LEM}
\end{figure}

\subsection{Impact of potential sources of \ion{O}{VII} emission}
\label{Sec:cooling_flow_cluster_core}

Due to the centrally peaked density profiles in relaxed galaxy clusters, the radiative cooling times of the diffuse gas trapped in the potential wells of these objects can become shorter than the Hubble time (typically less than $1$\,Gyr). It is now understood that the energy input from AGN feedback largely compensates for this gas cooling, keeping the ICM hot (for a review, see \citealp{2007ARA&A..45..117M}). However, the heating-cooling balance is rarely perfect, and local thermal instabilities can often develop, forming narrow strands of multi-phase gas (see e.g. \citealp{2012MNRAS.420.3174S, 2014ApJ...789..153L}). The coolest X-ray emitting phases of this gas should be visible through their soft X-ray line emission, for instance as \ion{O}{VII} or \ion{Fe}{XVII} \citep{2011MNRAS.412L..35S, 2014A&A...572L...8P}.

In order to investigate whether the potential X-IFU detection of this emission from cool gas would affect the analysis and results presented in this paper, we add an additional CIE component to our model. The typical temperatures for this cool CIE component as reported by \citet{2014A&A...572L...8P} are between $0.45-0.85$\,keV, while its normalisation is about $10-50$ times lower than the normalisation of the model describing the bulk of the intracluster medium. 

For our calculations we chose CE-$29$ \& A$383$ X-IFU observations (with exposure time $250$\,ks) as an example. We added an additional CIE component with temperature $0.6$\,keV, with a normalization that corresponds to a flux (in the $0.2-1$\,keV energy band) which is $10$ times lower than the flux of the original model as described in Sec.\,\ref{sec:observations_methods}. After simulating the new X-IFU spectra and fitting the spectra in the same manner as described in Sec.\,\ref{sec:summary:obs:methods}, the significance of the \ion{O}{VII} detection decreased from $8\sigma$ to $\approx 7.5\sigma$. 


Therefore we can conclude that, for the strongest absorbers like CE-$25$ and CE-$29$, even a $10$\% contribution in flux from the cooling gas would not matter, while for smaller $N_{\ion{O}{VII}}$ (like CE-$7$), caution should be taken with this cooling gas. If present, gas cooling will probably happen along patchy small-scale filaments that can be excluded from the spectral extraction to test the robustness of the \ion{O}{VII} absorption signal in case both phenomena are seen simultaneously. 

Furthermore, we estimated the effects of \ion{O}{VII} emission from gas belonging to the filament itself, by approximating the volume of the filament as a cylinder with a $100$\,kpc radius (same as our typical spectral extraction region), and a length of $2\times r_{200}$ (corresponding to one of the largest absorbers, i.e. CE-$29$ $z+$; see Fig.\,\ref{Fig:tau_vs_position}).  For a rough estimate, we convert the number of \ion{O}{VII} ions into a hydrogen number density by assuming an O/H metallicity of 0.2 Solar, and that all oxygen is in the \ion{O}{VII} ionization state, and averaging over all LOSs in the central $100$\,kpc. This yields an emission measure of the CE-$29$ $z+$ filament that is $4$ orders of magnitude lower than that of the potential cool gas in the cluster core, considered earlier in this section. We therefore conclude that the intrinsic \ion{O}{VII} emission from the filament does not affect the absorption signal.

\section{Conclusions}
\label{Sec:conclusions}
 
In this paper we studied the prospects of observing cosmic web filaments in absorption against diffuse X-ray emission of cool-core galaxy clusters. 


We extended the study of \citet{2021ExA....51.1043S}, which simplified the WHIM as a single temperature absorbing gas cloud in collisional ionisation equilibrium. We projected $23$ galaxy clusters from the Hydrangea cosmological hydrodynamical simulations along the $\pm x$, $\pm y$, and $\pm z$ simulation axes and out of the total number of $138$ directions we found $16$ directions with a column density sufficiently high that it could potentially be observed in \ion{O}{VII} absorption against galaxy clusters with the X-IFU instrument of the Athena X-ray Observatory (Fig.\ref{Fig:all_absorbers_histogram}, Fig.\,\ref{Fig:column_density_maps}, Fig.\,\ref{Fig:column_density_maps_app1} and Fig.\,\ref{Fig:column_density_maps_app2}).

We obtained the absorption profiles for all $16$ directions by averaging over $276$ sightlines in the simulation volumes, which were probing the area of a circle with radius $100$\,kpc in the plane of the sky. In Fig.\,\ref{Fig:tau_vs_position} and Fig.\,\ref{Fig:tau_vs_vtot_vs_energy} we showed the spatial properties and the velocity structure for all $16$ directions. 

For three absorbers (CE-$7$, CE-$25$, and CE-$29$ $z+$) we simulated the Athena X-IFU spectra, where all sightlines were weighted by the surface brightness profile of the chosen background galaxy clusters. Our simulations also took into account a three-component model for Galactic ISM absorption. The CE-$7$ galaxy cluster is at the low mass end of the galaxy cluster sample with mass $M_{200 \rm c} \approx  10^{14.34}$\,M$_{\astrosun}$, while the CE-$25$, and CE-$29$ $z+$ are among the three most massive Hydrangea clusters, with masses $M_{200 \rm c} \approx  10^{15.15}$\,M$_{\astrosun}$, and $M_{200 \rm c} \approx 10^{15.38}$\,M$_{\astrosun}$, respectively.

Our main results can be summarized as follows:
\begin{itemize}[leftmargin=5pt]
	\item Out of the massive clusters from the Hydrangea cosmological simulations, $16$ out of $138$ directions have on average \ion{O}{VII} column densities above $10^{14.5}$\,cm$^{-2}$.  The strongest of these absorbers can be detected with at least 5$\sigma$ significance in reasonable exposure times with the Athena X-IFU ($T_{\rm{exp}} \leq 250$\,ks), provided that the Galactic foreground absorption at the observed line energy permits it. Table \ref{Tab:fitting_results_exposure} summarizes the exposure times for \ion{O}{VII} detection for three example absorbers: CE-$7$ $x-$, CE-$25$ $x+$, and CE-$29$ $z+$. Unfortunately, \ion{O}{VIII} absorption from the WHIM cannot be detected with Athena since the wings of the \ion{O}{VIII} galaxy cluster emission line are too bright for \ion{O}{VIII} absorption line to be sufficiently separated with the X-IFU resolution of $2.5$\,eV. 
	
	\item Under the assumption that the ISM of the Milky Way can be represented by a three-component model as described in Sec.\,\ref{Sec:Galaxy_absorption}, the ISM lines that might intervene with the \ion{O}{VII} cosmic web filaments detection are \ion{N}{VII}, \ion{O}{I}, \ion{O}{II}, \ion{O}{III}, and \ion{O}{VII} (Fig.\,\ref{Fig:CE_25_all_abell_clusters}).
	
	\item From our studies, the most promising clusters for the detection of the \ion{O}{VII} WHIM absorption against extended sources seem to be the higher redshift clusters, e.g. A2390 ($z=0.2302$) and A383 ($z=0.1883$), for which all three Hydrangea absorbers with different total velocity along the line of sight  $v_{\rm TOT}$ and different optical depths could be detected with X-IFU. Lower redshift clusters give even more significant detections, but suffer more from the presence of foreground ISM lines, and the detection or non-detection depends on the combination of the galaxy cluster redshift and the velocity $v_{\rm TOT}$. 
	
	
	\item The velocity difference between the strongest absorbers and the galaxy cluster centre is on average $\sim \pm 1000$\,km/s (Fig.\,\ref{Fig:tau_vs_vtot_vs_energy}). The maximum $v_{\rm TOT}$ that we found in the Hydrangea absorbers was $\approx 2200$\,km/s for CE-$28$. 
	
	
	
	\item Even though the larger spectral extraction region with $R = 300$\,kpc results in shallower absorption profiles in comparison with $R = 40$\,kpc, the increase in the photon number count for the larger area improves the significance of the \ion{O}{VII} detection (Fig.\,\ref{Fig:quasar_like_studies_compar}).
	
	 
	
	
	\item If the spectral resolution of LEM is approximately $1$\,eV for extended sources, then it can additionally probe the WHIM \ion{O}{VII} line which is closer in energy to Galactic foreground lines (see Fig.\,\ref{Fig:spectrum_A2029_CE25_LEM}). The unprecedented spectral resolution of LEM could even distinguish the \ion{O}{VIII} emission line from the galaxy cluster and the \ion{O}{VIII} absorption line from the WHIM. According to our simulations, a $5\sigma$ significance of the \ion{O}{VIII} WHIM detection for cases like CE-$25$ \& A$2029$ would be feasible in $250$\,ks (see Sec.\,\ref{Sec:discussion_detectors}). 
  
	\item  If the Athena X-IFU resolution was degraded to $4$\,eV, cases similar to \ion{O}{VII} absorption from CE-$25$ and the A$2029$ background galaxy cluster would be lost due to blending with ISM absorption lines. For CE-$25$ and background clusters A$1413$ and A$262$ this would result in an increase of the required exposure time from  $93$\,ks to $174$\,ks and from $25$\,ks to $63$\,ks for a detection of at least 5$\sigma$, respectively. 
\end{itemize}

\section*{Acknowledgements}
The authors acknowledge the financial support from NOVA, the Netherlands Research School for Astronomy. L.{\v S}. is supported by NWO grant Athena 184.034.002. A.S. is supported by the Women In Science Excel (WISE) programme of the Netherlands Organisation for Scientific Research (NWO), and acknowledges the Kavli IPMU for the continued hospitality. SRON Netherlands Institute for Space Research is supported financially by NWO. N.A.W. is supported by a CIERA Postdoctoral Fellowship. Y.B. acknowledges funding from the Dutch Research Organisation (NWO) through Veni grant number 639.041.751 and financial support from the Swiss National Science Foundation (SNSF) under project 200021\_213076. This work used the DiRAC@Durham facility managed by the Institute for Computational Cosmology on behalf of the STFC DiRAC HPC Facility (www.dirac.ac.uk). The equipment was funded by BEIS capital funding via STFC capital grants ST/K00042X/1, ST/P002293/1 and ST/R002371/1, Durham University and STFC operations grant ST/S003908/1. DiRAC is part of the National e-Infrastructure.

\section*{Data Availability}
The dataset generated and analysed during this study is available in the ZENODO repository  \citep{lydia_stofanova_2023_reproduction}.




\bibliographystyle{mnras}
\bibliography{bibliography} 


\appendix

\section{Total column density maps}
\label{Sec:appendix_column_density_maps}
\begin{figure*}
	\centering
	\includegraphics[width=0.4\textwidth]{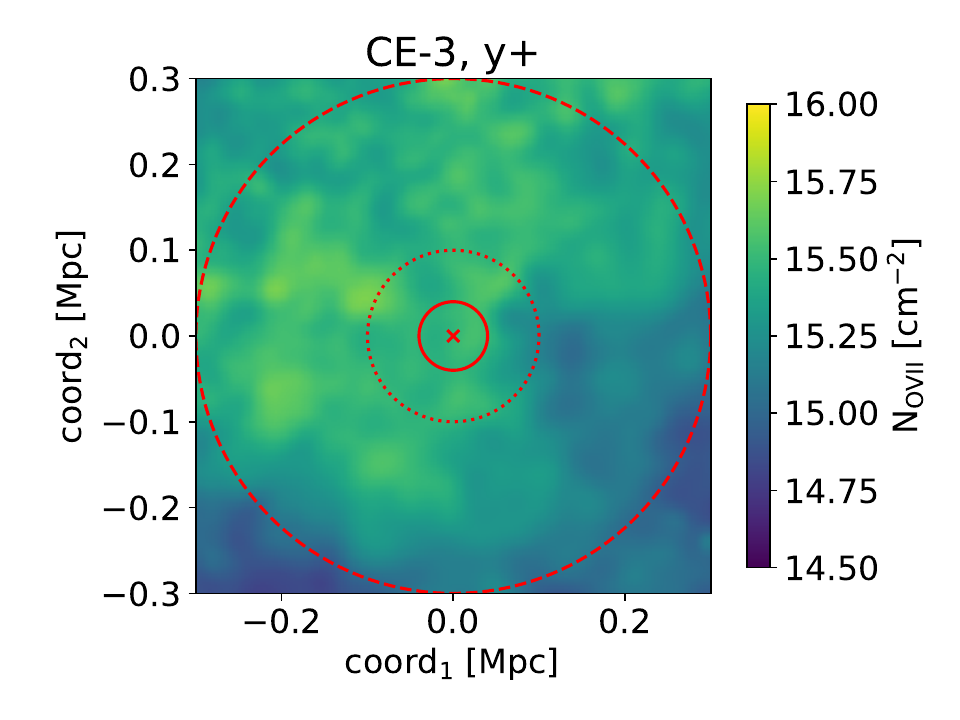} 
	\includegraphics[width=0.4\textwidth]{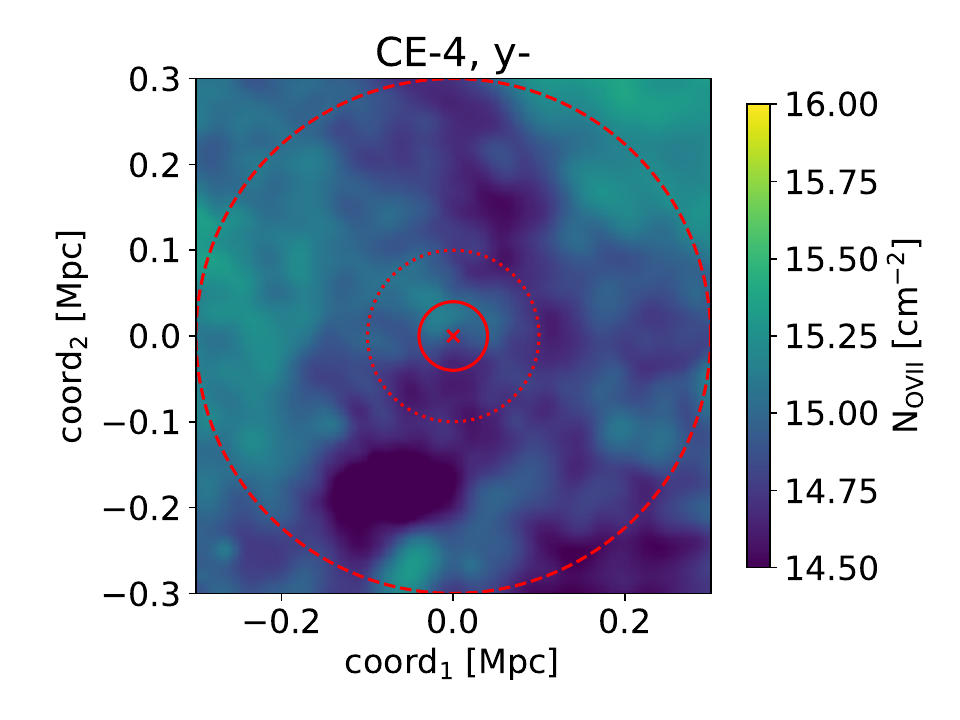} \\
	\includegraphics[width=0.4\textwidth]{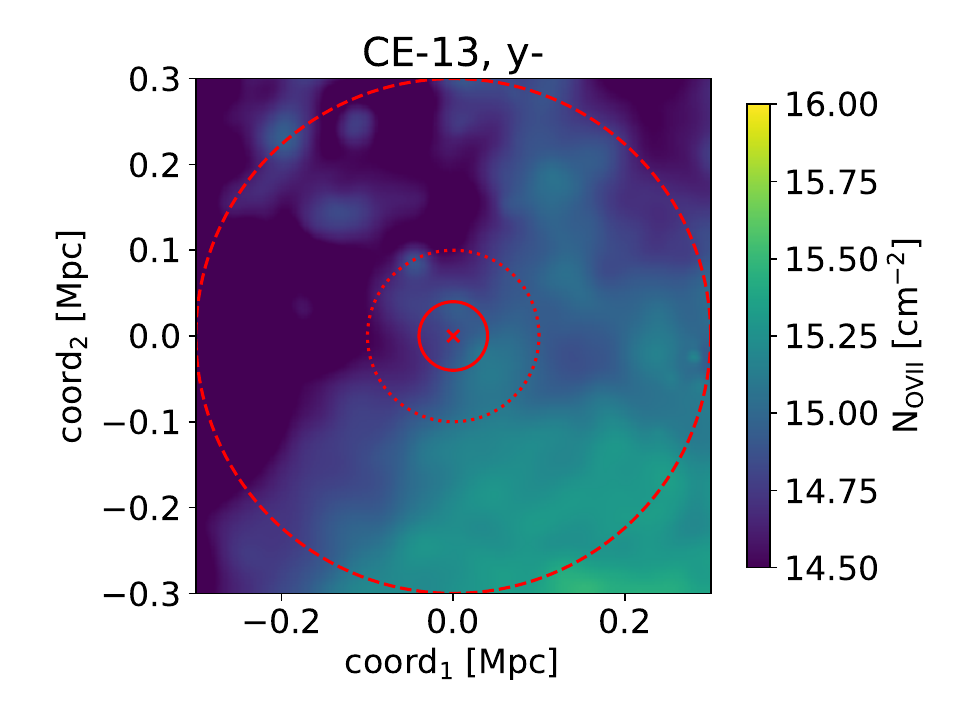} 
	\includegraphics[width=0.4\textwidth]{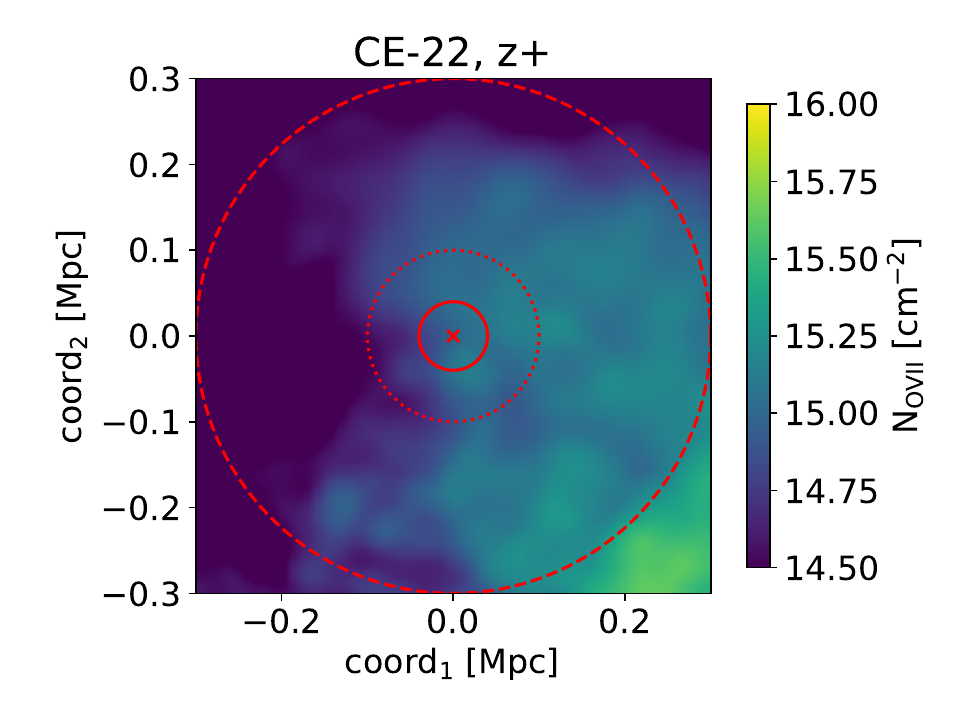} \\
	\includegraphics[width=0.4\textwidth]{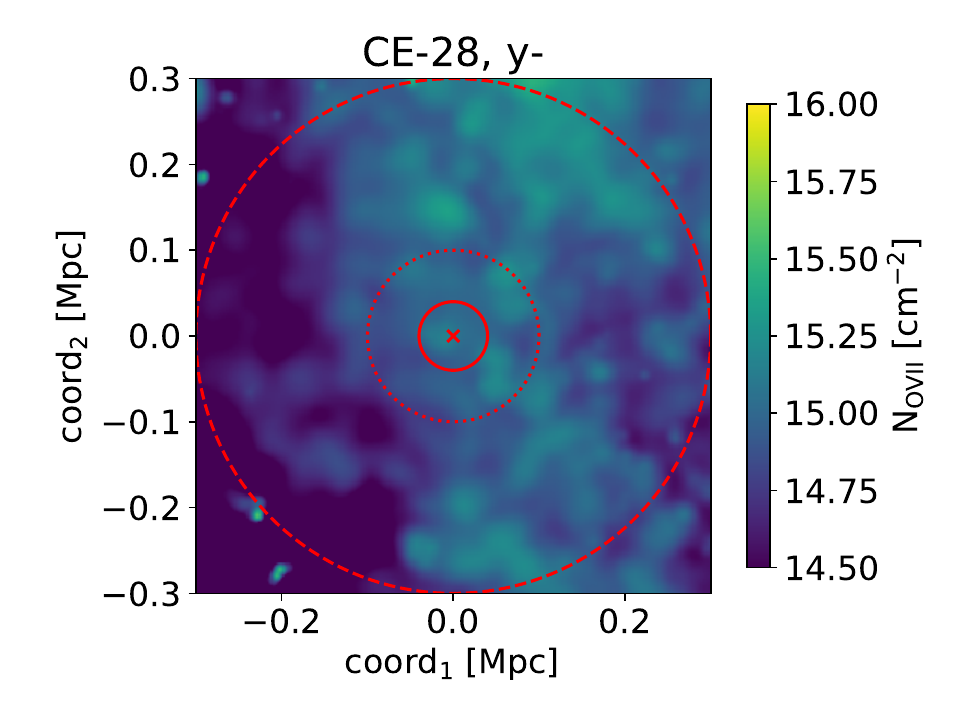} 
	\includegraphics[width=0.4\textwidth]{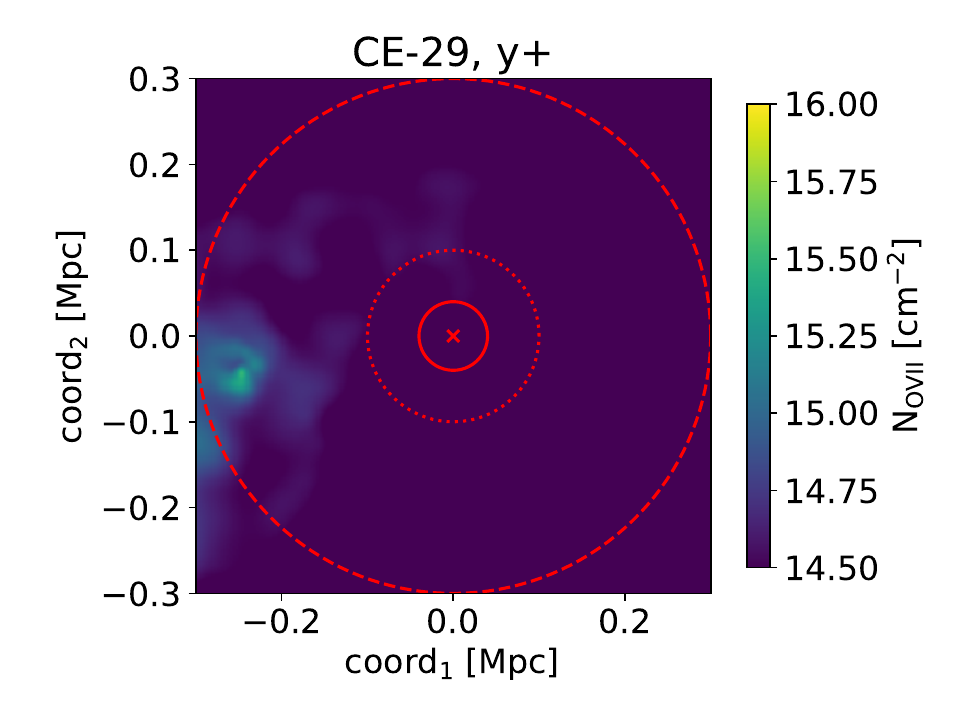} \\
	\caption{Continuation of Fig.\,\ref{Fig:column_density_maps} for CE-$3$ $y+$, CE-$4$ $y-$, CE-$13$ $y-$, CE-$22$ $z+$, CE-$28$ $y-$, and CE-$29$ $y+$.}	
	\label{Fig:column_density_maps_app1}
\end{figure*}

\begin{figure*}
	\centering
	\includegraphics[width=0.4\textwidth]{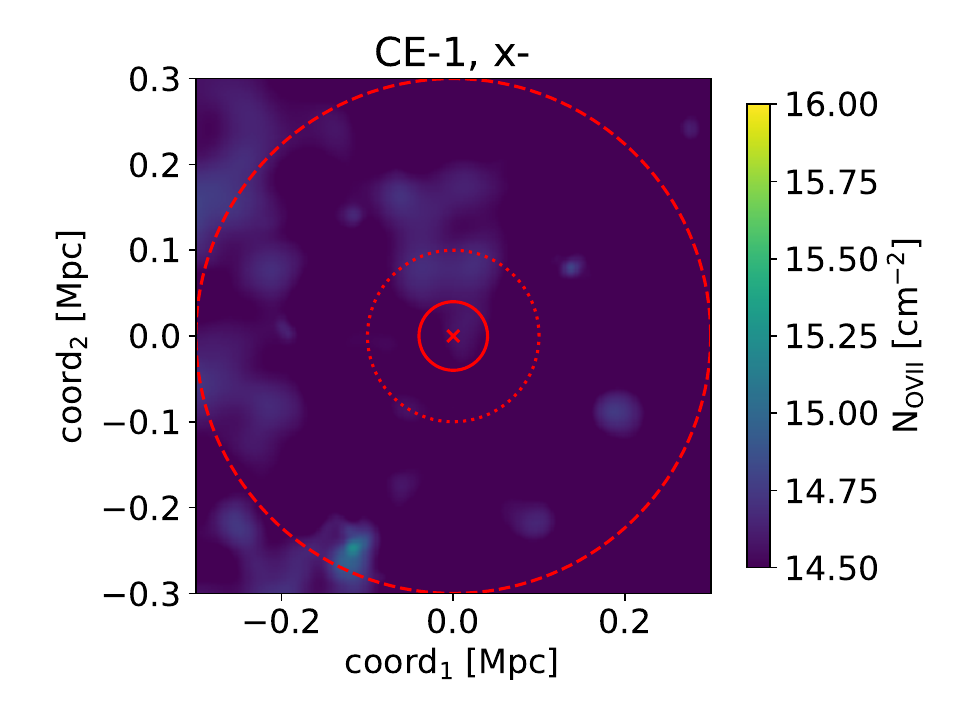} 
	\includegraphics[width=0.4\textwidth]{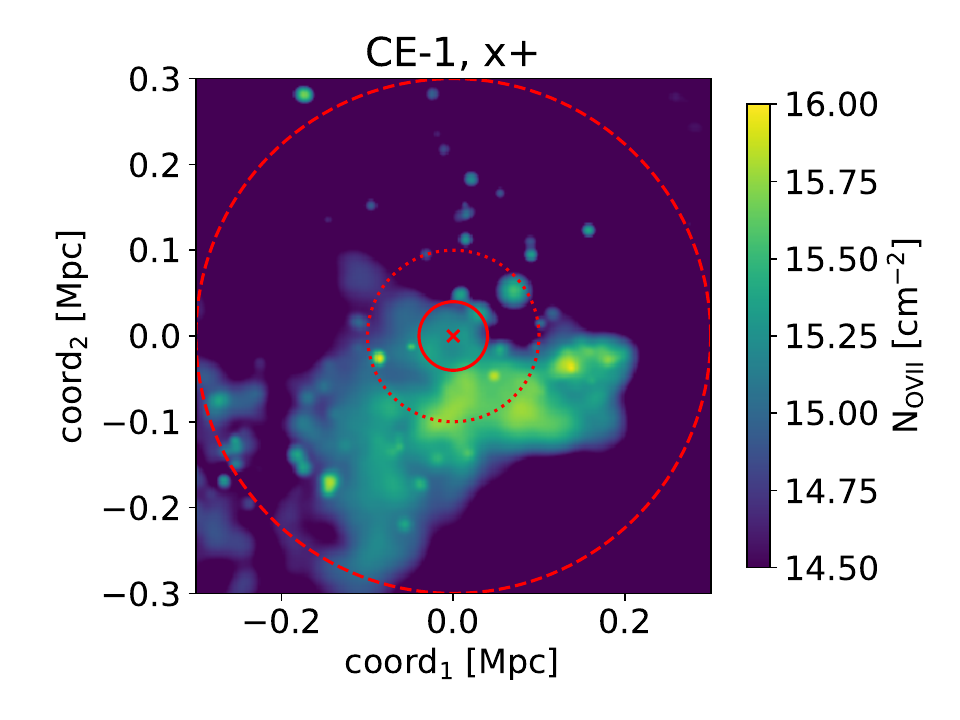} \\
	\includegraphics[width=0.4\textwidth]{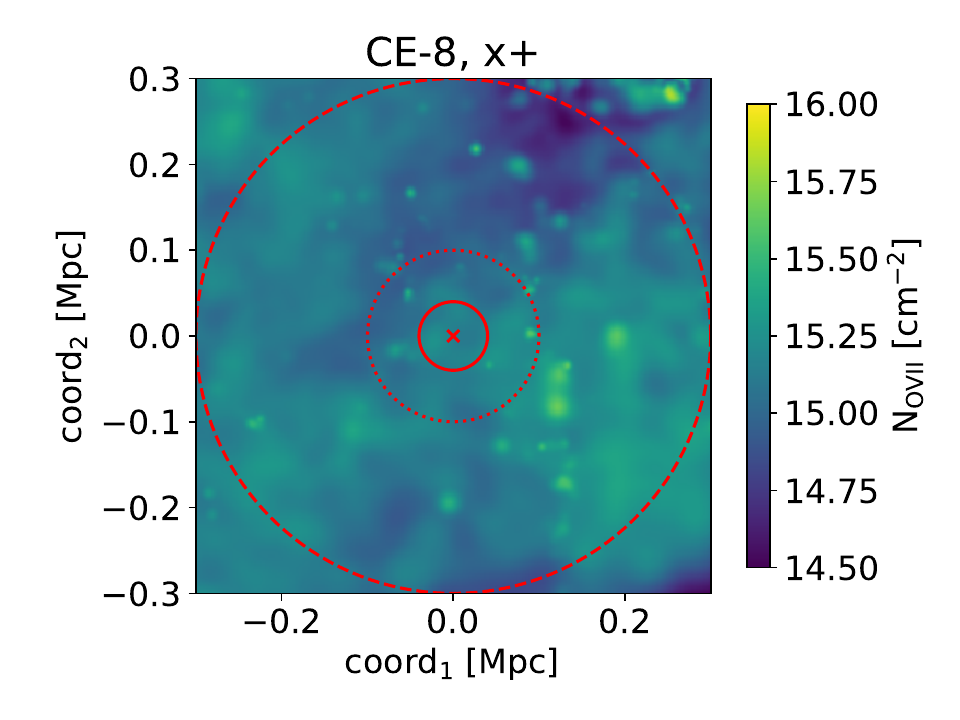} 
	\includegraphics[width=0.4\textwidth]{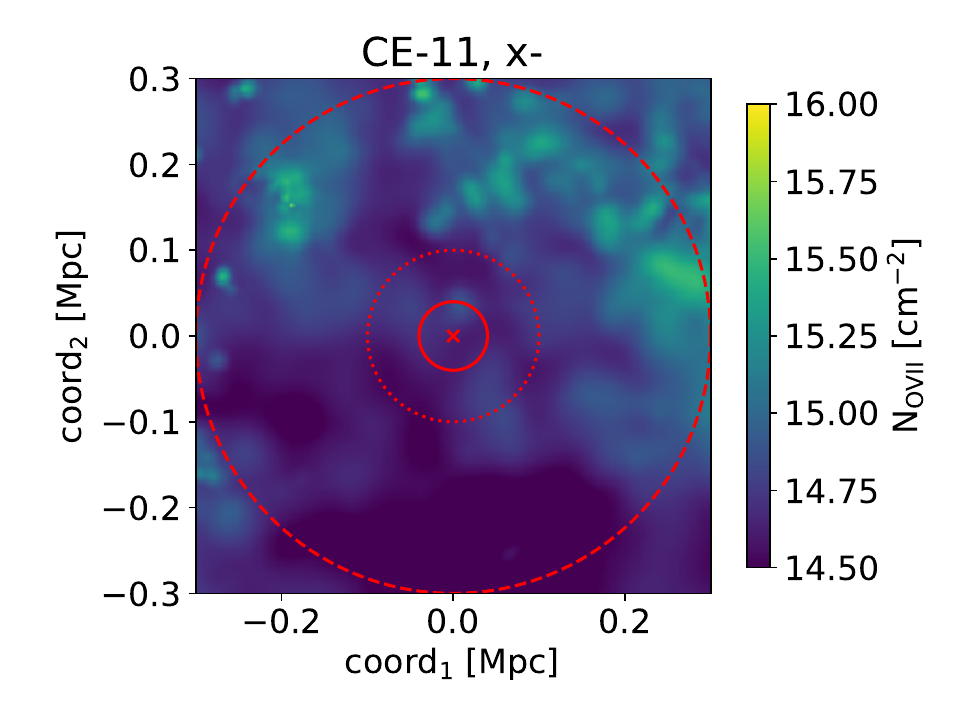} \\ 
	\includegraphics[width=0.4\textwidth]{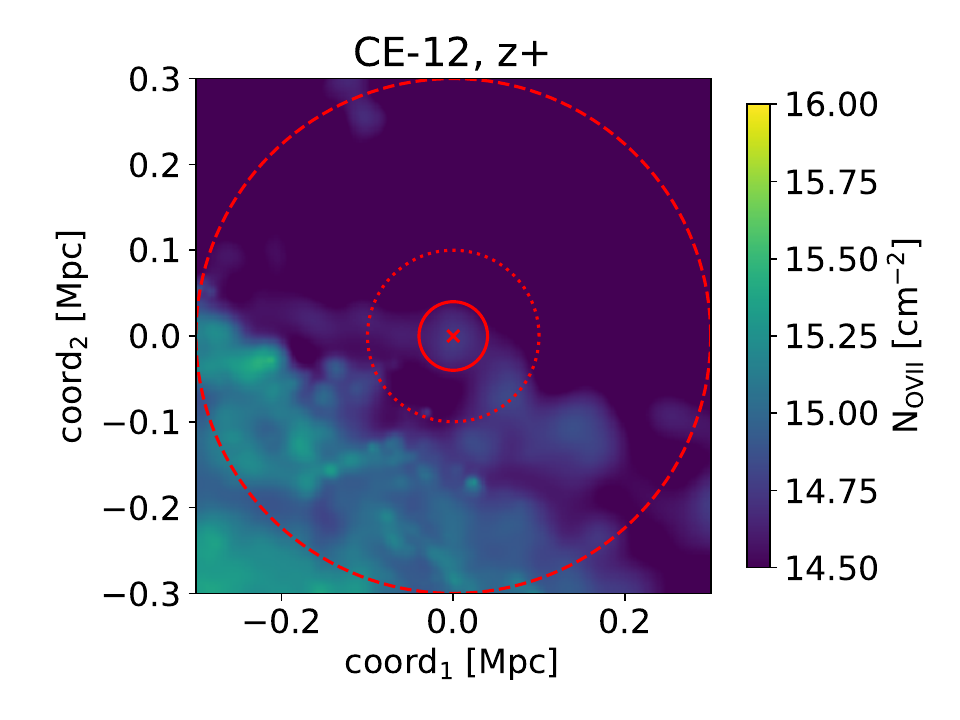} 
	\includegraphics[width=0.4\textwidth]{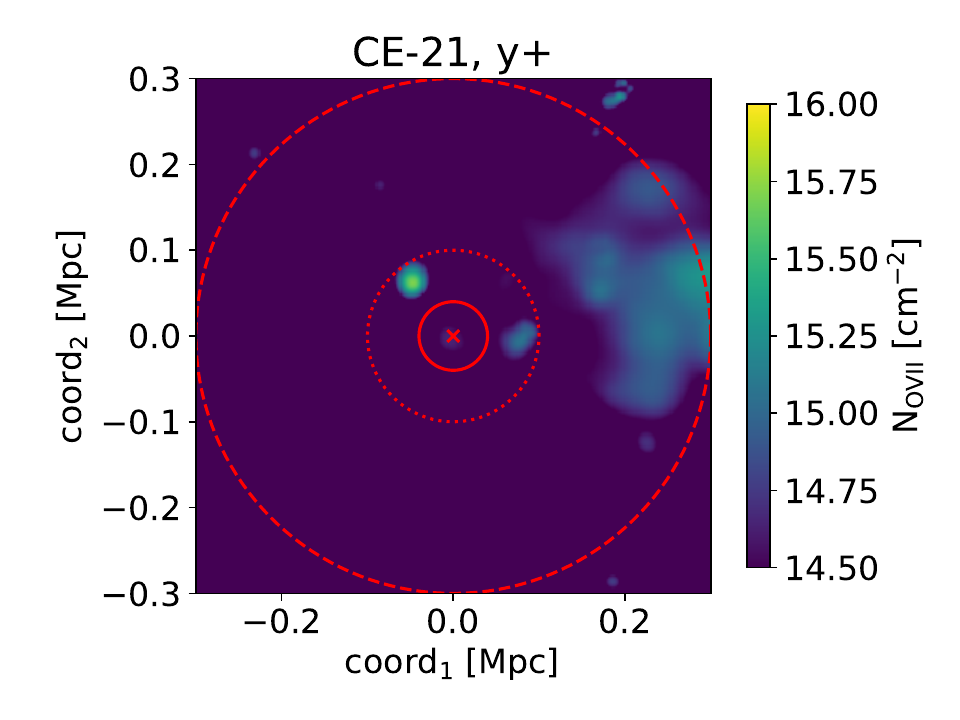} \\
	\includegraphics[width=0.4\textwidth]{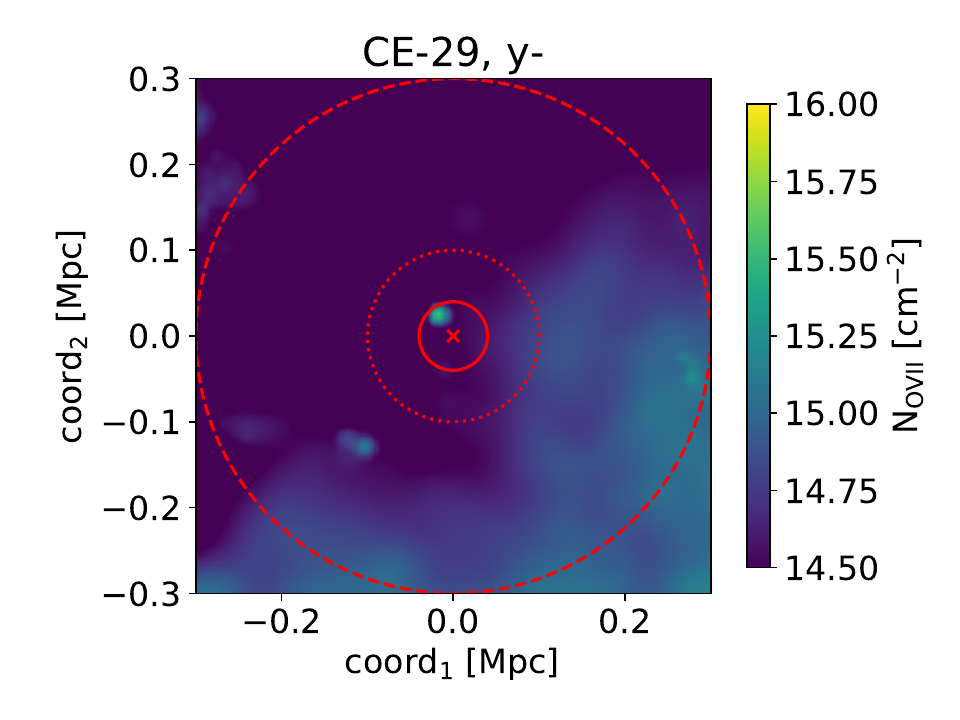} \\
	\caption{Continuation of Fig.\,\ref{Fig:column_density_maps} for CE-$1$ $x-$, CE-$1$ $x+$, CE-$8$ $x+$, CE-$11$ $x-$, CE-$12$ $z+$, CE-$21$ $y+$, and CE-$29$ $y-$.}	
	\label{Fig:column_density_maps_app2}
\end{figure*}


\bsp	
\label{lastpage}
\end{document}